\documentclass[jgrga]{agu2001}
\usepackage{graphicx}
\authorrunninghead{mori ET AL.}

\titlerunninghead{CONTINUUM LIMIT OF BURRIDGE-KNOPOFF MODEL}

\authoraddr{H. Kawamura,
Department of Earth and Space Science, Faculty of Science,
Osaka University, Toyonaka 560-0043,
Japan. (kawamura@ess.sci.osaka-u.ac.jp)}

\begin{document}

%
%
%
%
%

%
%

\title{Spatiotemporal correlations  of earthquakes in the continuum limit of the one-dimensional Burridge-Knopoff model}
%

%
%


\author{Takahiro Mori and Hikaru Kawamura}
\affil{Department of Earth and Space Science, Faculty of Science,
Osaka University, Toyonaka 560-0043,
Japan}

\begin{abstract}
Spatiotemporal correlations of the one-dimensional spring-block (Burridge-Knopoff) model of earthquakes, either with or without the viscosity term, are studied by means of numerical computer simulations. The continuum limit of the model is examined by systematically investigating the model properties with varying the block-size parameter $a$ toward $a\rightarrow 0$. The Kelvin viscosity term is introduced so that the model dynamics possesses a sensible continuum limit. In the presence of the viscosity term, many of the properties of the original discrete BK model are kept qualitatively unchanged even in the continuum limit, although the size of minimum earthquake gets smaller as $a$ gets smaller.  One notable exception is the existence/non-existence of the doughnut-like quiescence prior to the mainshock.  Although large events of the original discrete BK model accompany  seismic acceleration  together with a doughnut-like quiescence just before the mainshock, the spatial range of the doughnut-like quiescence becomes narrower as $a$ gets smaller,  and in the continuum limit, the doughnut-like quiescence  might vanish altogether. The doughnut-like quiescence observed in the  discrete BK model is then a phenomenon closely related to the short-length  cut-off scale of the model.
\end{abstract}

%
%

%

\begin{article}

%
%

\section{Introduction}

 An earthquake is a stick-slip dynamical instability of a pre-existing fault driven by the motion of a tectonic plate [\textit{Scholz}, 1998, 2002]. While an earthquake is a complex phenomenon, certain empirical laws such as the Gutenberg-Richter (GR) law and the Omori law concerning its statistical properties are known to hold. One fruitful strategy in understanding such statistical properties of earthquakes might be to properly model an earthquake fault and to elucidate the model properties by numerical computer simulations.  One of popular statistical models of an earthquake fault is the so-called spring-block model originally proposed  by Burridge and Knopoff (BK) [\textit{Burridge and Knopoff}, 1967]. In this model, an earthquake fault is simulated by an assembly  of blocks, each of which is connected via the elastic springs to the  neighboring blocks and to the moving plate. All blocks are subject to  the friction force, the source of the nonlinearity in the model,  which eventually realizes an earthquake-like frictional  instability. While the spring-block model is obviously a crude model  to represent a real earthquake fault, its simplicity enables one to  study the statistical properties with high precision.

 Carlson, Langer and others [\textit{Carlson and Langer}, 1989a;  \textit{Carlson and Langer}, 1989b; \textit{Carlson et al.}, 1991;  \textit{Carlson}, 1991a; \textit{Carlson}, 1991b; \textit{Shaw et  al.}, 1992; \textit{Carlson et  al.}, 1994; \textit{Schmittbuhl et al.}, 1996]  studied the statistical properties of the BK model quite  extensively both in one dimension (1D) and two dimensions (2D), paying particular attention to the magnitude distribution of earthquake events. In these studies, a simple velocity-weakening friction law was used as a constitutive law. The spring-block model has also  been extended in several ways, {\it e.g.\/}, taking account of the  effect of the viscosity [\textit{Myers and Langer}, 1993;  \textit{Shaw}, 1994; \textit{De and Ananthakrisna}, 2004], modifying  the form of the friction force [\textit{Myers and Langer}, 1993;  \textit{Shaw}, 1995; \textit{De and Ananthakrisna}, 2004],  considering the long-range interactions between blocks [\textit{Xia  et al.}, 2005, 2008a; \textit{Mori and Kawamura}, 2008b], driving the system only at one end of the  system [\textit{Vieira}, 1992], or by incorporating the rate- and  state-dependent friction law [\textit{Ohmura and Kawamura}, 2007].

 In the previous papers, the present authors also performed an extensive numerical study of the statistical  properties of the 1D and 2D BK models with the velocity-weakening friction law for a wide parameter range, focusing on their {\it spatiotemporal  correlations\/} [\textit{Mori and Kawamura}, 2005; 2006; 2008a; Kawamura, 2006]. These studies have revealed several interesting features of the BK model. For example, the model exhibits either ``near-critical'', ''supercritical'' or ``subcritical'' behaviors primarily depending on its frictional parameter representing the extent of the velocity weakening. Preceding the mainshock, the  frequency of smaller events is gradually enhanced, whereas, just before the mainshock, it is  suppressed in a close vicinity of the epicenter of the upcoming event (the Mogi doughnut). The apparent $B$-value of the magnitude  distribution changes significantly preceding the mainshock. 

 Although the BK model has widely been used as a useful tool to investigate statistical properties of earthquakes, the block discretization inherent to the model construction is a crude approximation of the originally continuum earthquake fault. It introduces the short-length cut-off scale into the problem. Therefore, in order to check the validity of the model, it is crucially important to examine the continuum limit of the BK model. Indeed, Rice criticized that the discrete BK model with the velocity-weakening friction law was ``intrinsically discrete'', lacking  in a well-defined continuum limit [\textit{Rice}, 1993]. Rice argued that the spatiotemporal complexity observed in the discrete BK model was due to the inherent discreteness of the model, which should disappear in continuum. This problem of the continuum limit of the BK model was also discussed by Myers and Langer [\textit{Myers and Langer}, 1993], who introduced the Kelvin viscosity term to produce a small length scale which allows a well-defined continuum limit. Myers and Langer, and subsequently Shaw [\textit{Shaw}, 1994], observed that the added viscosity term smoothed the rupture dynamics, apparently giving rise to the continuum limit accompanied by the spatiotemporal complexity.

 The BK model represents an earthquake fault by an assembly of blocks and springs. Conversely, one can regard the BK model as obtained by discretizing an appropriate wave equation describing the continuum. In one spatial dimension, an appropriate wave equation might be given by the partial differential equation, 
\begin{equation}
\begin{array}{ll}
\frac{\partial^2 U(x^{\prime},t^{\prime})} {\partial {t^{\prime}}^2}
=\omega^2(\nu^{\prime} t^{\prime}-U(x^{\prime},t^{\prime}))
\ \ \ \ \ \ \ \ \ \ \ \ \ \\ \ \ \ \ \ \ \ \ \ \ \ \ \
+{\xi^{\prime}}^2 
\frac{\partial^2 U(x^{\prime},t^{\prime})}{\partial
{x^{\prime}}^2}-\phi^{\prime}(\frac{\partial
U(x^{\prime},t^{\prime})} {\partial t^{\prime}}),
\end{array}
\end{equation}
where $U(x^{\prime},t^{\prime})$ is the displacement at the spatial position $x^{\prime}$ and time $t^{\prime}$, $\nu^{\prime}$  the loading rate by the plate motion, $\xi^{\prime}$  the wave velocity, $\omega$ the characteristic frequency 
and $\phi^{\prime}$ is the friction force per unit mass. 
One may discretize Eq.(1) as
\begin{equation}
\begin{array}{ll}
\frac{\partial^2 U_i} {\partial {t^{\prime}}^2}
=\omega^2(\nu^{\prime} t^{\prime}-U_i)
\ \ \ \ \ \ \ \ \ \ \ \ \ \\ \ \ \ \ \ \ \ \ \ \ \ \ \
+\frac{{\xi^{\prime}}^2}{{a^{\prime}}^2}(U_{i+1}-2U_i+U_{i-1})
-\phi^{\prime}(\frac{\partial U_i} {\partial t^{\prime}}),
\end{array}
\end{equation}
where $a^{\prime}$ is the grid spacing corresponding to the block size of the BK model.  If one regards $U_i$ as the displacement of the $i$-th block, Eq.(2) is the equation of motion of the 1D BK model. Conversely, if one takes the limit $a^{\prime}\rightarrow 0$ starting from the discrete BK model (2), one can obtain the continuum limit of the BK model.

 As mentioned, the naive continuum limit of the discrete BK model defined by Eq.(2) has a problem in that the pulse of slip tends to become increasingly narrow in width in the limit $a^{\prime} \to 0$,  {\it i.e.}, the dynamics becomes sensitive to the grid spacing $a^{\prime}$. One way to circumvent this problem is to introduce the viscosity term $\eta^{\prime} \partial^3 U/\partial {x^{\prime}}^2 \partial t^{\prime}$ into Eq.(2) to produce a small length scale. Myers and Langer showed that, owing to the added viscosity term, the system became independent of the grid spacing $a^{\prime}$ as long as a new small length scale $\epsilon^{\prime}$, defined by  
\begin{equation}
\epsilon^{\prime} =\pi
\sqrt{\frac{\eta^{\prime}}{\alpha \omega}},
\end{equation}
is sufficiently larger than the grid spacing $a^{\prime}$ [\textit{Myers and Langer}, 1993]. Here, $\eta^{\prime}$ is a viscosity coefficient, and $\alpha$ is the velocity-weakening parameter representing the rate of the friction force getting weaker on increasing the sliding velocity (to be defined below in Eq.(7)). Shaw also showed by adding the viscosity term to the 1D BK model that the magnitude distribution became independent of the grid spacing $a^{\prime}$ for sufficiently small $a^{\prime}$ [\textit{Shaw}, 1994]. 

 In the present paper, following Shaw, we numerically examine the continuum limit of the 1D BK model by performing extensive numerical simulations on the BK model with successively smaller grid spacings $a^{\prime}$. Namely, with varying the grid spacing $a^{\prime}$, we study  how seismic spatiotemporal correlations of the 1D BK model vary and approach the continuum limit. The two types of the BK model, {\it i.e.\/}, the one without the viscosity term (the original BK model) and the one with the viscosity term (the modified BK model) are studied. Our work can be regarded as an extension of the work of Shaw concerning the magnitude distribution of the 1D BK model to various types of seismic spatiotemporal correlations of the same model.

 The present paper is organized as follows. In \S 2, we introduce the model and explain some of the details of our numerical simulation. In \S 3, we examine the continuum limit of the 1D BK model without the viscosity term. We analyze the magnitude distribution and various types of seismic spatiotemporal correlation functions, including the recurrence-time distribution, the time-correlation function of seismic events before and after the mainshock, the time development of the spatial correlation function of seismic events before the mainshock, and the time development of the magnitude distribution function  before and after the mainshock. In \S 4, we make a similar analysis for the 1D BK model with the viscosity term. Finally, \S 5 is devoted to summary and discussion.

%
%


%
%

\section{The model and the method}

Let us begin with the 1D wave equation in the continuum with the Kelvin viscosity term $\eta^{\prime} \partial^3 U/\partial {x^{\prime}}^2 \partial t^{\prime}$,
\begin{equation}
\frac{\partial^2 U} {\partial {t^{\prime}}^2}
=\omega^2(\nu^{\prime} t^{\prime}-U)
+{\xi^{\prime}}^2\frac{\partial^2 U}{\partial {x^{\prime}}^2}
+\eta^{\prime}\frac{\partial^3 U}{\partial {x^{\prime}}^2 \partial t^{\prime}}
-\phi^{\prime}(\frac{\partial U_i} {\partial t^{\prime}}).
\end{equation}
In order to make the equation dimensionless, we measure the time $t'$ in units of the characteristic frequency $\omega$, the coordinate $x^{\prime}$ in units of $\xi^{\prime}/\omega$ and the displacement $U_i$ in units of the length $\mathcal{L}=\phi^{\prime}(0)/\omega^2$, $\phi^{\prime}(0)$ being a static friction per unit mass. Then, the equation of motion can be written in the dimensionless form as
\begin{equation}
\frac{\partial^2 u} {\partial t^2}=\nu t-u
+\frac{\partial^2 u}{\partial x^2}
+\eta\frac{\partial^3 u}{\partial x^2 \partial t}
-\phi(\frac{\partial u} {\partial t}),
\end{equation}
where $t=t^{\prime}\omega$ is the dimensionless time, $x=x^{\prime}/(\xi^{\prime}/\omega)$ the dimensionless coordinate, $u=U/\mathcal{L}$ the dimensionless displacement, $\phi(\dot{u})=\phi^{\prime}(\dot{U})/(\mathcal{L}\omega^2)$ the dimensionless friction force, and $\eta=\eta^{\prime}/({\xi^{\prime}}^2/\omega)$ is the dimensionless viscosity coefficient.

After the block discretization, one gets
\begin{equation}
\begin{array}{ll}
\ddot u_i=\nu t-u_i+\frac{1}{a^2}(u_{i+1}-2u_i+u_{i-1})
\ \ \ \ \ \ \ \ \ \ \ \ \ \\ \ \ \ \ \ \ \ \ \ \ \ \ \
+\frac{\eta}{a^2}(\dot{u}_{i+1}-2\dot{u}_i+\dot{u}_{i-1})
-\phi (\dot u_i),
\end{array}
\end{equation}
where $a=a^{\prime}/(\xi^{\prime}/\omega)$ is the dimensionless grid spacing, and the notation $\dot{u}=\partial u / \partial t$ has been used. If we put $\eta=0$ in Eq.(6), it reduces to the BK model used in the previous papers with the correspondence $l=1/a$ [{\textit Mori and Kawamura}, 2005; 2006]. Note that the dependence on $\xi^{\prime}$ has been adsorbed into the definition of the dimensionless variables. In the following, we study the statistical properties of the model (6) with varying the dimensionless grid spacing $a$ in order to see how the spatiotemporal correlation functions behave in the continuum limit $a \to 0$. We investigate both cases of the non-viscosity model with $\eta=0$ and the viscosity model with $\eta=0.02$. 

 In order for the model to exhibit a dynamical instability corresponding to an earthquake, it is essential that the friction force $\phi$ possesses a frictional {\it weakening\/} property, {\it i.e.\/}, the friction should become weaker as the block slides. Here, we assume the velocity-weakening friction force. The velocity-weakening friction force used by Carlson [\textit{Carlson et al.}, 1991]  was defined by
\begin{equation}
\phi(\dot u) = \left\{ 
             \begin{array}{ll} 
             (-\infty, 1],  & \ \ \ \ {\rm for}\ \  \dot u_i\leq 0, \\ 
              \frac{1-\sigma}{1+2\alpha \dot u_i/(1-\sigma )}, &
             \ \ \ \ {\rm for}\ \  \dot u_i>0, 
             \end{array}
\right.
\end{equation}
where its maximum value corresponding to the static friction has been normalized to unity. As noted above, this normalization condition $\phi(\dot u=0)=1$ has been utilized to set the length unit  $\mathcal{L}$. The back-slip is inhibited by imposing an infinitely large friction for $\dot u_i<0$, {\it i.e.\/}, $\phi(\dot u<0)=-\infty $. The friction force is characterized by the two parameters, $\sigma$ and $\alpha$. The former, $\sigma$, introduced by [{\textit Carlson et al.}, 1991] as a technical device,  represents an instantaneous drop of the friction force at the onset of the slip, while the latter, $\alpha$, represents the rate of the friction force getting weaker on increasing the sliding velocity. In the present simulation, we regard $\sigma$ to be small and fix $\sigma =0.01$.

 Previous studies on the model has revealed that the velocity-weakening parameter $\alpha$ is most important among various parameters of the model in determining its statistical properties [\textit{Mori and Kawamura}, 2005; 2006; 2008a; 2008b].  In the present study, we set $\alpha$ either $\alpha=1$ or $\alpha=3$ as two typical cases. The former  $\alpha=1$ corresponds to the ``near-critical'' case in which the magnitude distribution exhibits a near straight-line behavior close to the GR law, whereas the latter case $\alpha=3$ corresponds to the ``supercritical'' case in which the magnitude distribution exhibits  a characteristic peak at a larger magnitude with the GR-law-like behavior realized only at smaller magnitudes. 

 We solve the equation of motion (6) by using the Runge-Kutta method of the fourth order. The width of the time discretization $\Delta t$ is taken to be $\Delta t =10^{-3}$. Total number of $10^6\sim10^7$ events are generated in each run, which are used to perform various averagings. In calculating the observables, initial $10^4\sim 10^5$ events are discarded as transients. In order to eliminate the possible finite-size effects, the total number of blocks $N$ are taken to be large, $N=L/a$ with fixing $L=200$, periodic boundary condition being applied. In a very special occasion where the finite-size effect is most severe, we simulate even larger system of $L=400$. The continuum limit of the model is investigated by varying the dimensionless grid spacing $a$ in the range $1 \geq a \geq 1/32$. The total CPU time is about 3,600 hours of of Pentium D.

 The dimensionless distance $r$ between the block $i$ and $i^{\prime}$ is measured by
\begin{equation}
r=a|i-i^{\prime}|.
\end{equation}
The small length scale $\epsilon^{\prime}$ defined by Eq.(3) is also given in the dimensionless form as 
\begin{equation}
\epsilon=\epsilon^{\prime}/(\xi^{\prime}/\omega)
=\pi \sqrt{\frac{\eta}{\alpha}}.
\end{equation}
The magnitude of a seismic event, $\mu$, is defined by a logarithm of its moment $M$, {\it i.e.\/},
\begin{equation}
\mu= \ln M = \ln \left( \sum_i a \Delta u_i \right),
\end{equation}
where $\Delta u_i$ is the total displacement of the $i$-th block
during a given event and the sum is taken over all blocks involved in
the event [\textit{Carlson et al.}, 1991].

\section{The non-viscosity model}

 In this section, we analyze the continuum limit of the 1D {\it non-viscosity} BK model with $\eta=0$. We show the results of our numerical simulations for various observables, including the recurrence-time distribution, the time-correlation function of seismic events before and after the mainshock, the time development of the spatial correlation function of seismic events before the mainshock, and the time development of the magnitude distribution function  before and after the mainshock, with varying the block-size parameter $a$ in the range $1\geq a\geq 1/32$. We set $\alpha$ either $\alpha=1$ or $\alpha=3$. As mentioned,  $\alpha=1$ corresponds to the ``near-critical'' case while  the latter  $\alpha=3$ corresponds to the ``supercritical'' case.

\subsection{The magnitude distribution}

 The magnitude distributions $R(\mu)$ of the 1D non-viscosity BK model is shown in Figs. 1(a) and (b) for the cases of  $\alpha=1$ and 3, respectively.  The grid spacing  $a$ is varied in the range $1 \geq a \geq 1/32$.   One can see from the figures that the qualitative features of $R(\mu)$, {\it i.e.\/}, a ``near-critical'' feature in the case of $\alpha=1$ and a ``supercritical'' feature in the case of $\alpha=3$, are preserved even in the continuum limit. Meanwhile, in both cases of $\alpha=1$ and $\alpha=3$, the magnitude distribution is considerably widened as the grid spacing $a$ becomes smaller. The maximum magnitude gets bigger and and the minimum magnitude gets smaller. The observation that the size of minimum event gets smaller for smaller $a$ can naturally be understood, since $a$ sets a small cut-off length scale.  As can be clearly seen from the figures, $R(\mu)$ does not seem to converge to an asymptotic form even at our smallest grid spacing $a=1/32$. Namely, $R(\mu)$ continues to change its form as $a$ gets smaller, although the change seems to be limited to that of the range and the extent of $R(\mu)$ for $a\leq 1/8$. Similar results were reported by [\textit{Shaw}, 1994].


\subsection{The local recurrence-time distribution}

 Next, we examine the continuum limit of the recurrence-time distribution of the model. Following our previous studies on the discrete BK model [\textit{Mori and Kawamura}, 2005; 2006], we study here the {\it local\/} recurrence-time distribution function  $P(T)$. The local recurrence time $T$ is defined by the time until the  next event occurs with its epicenter lying in a vicinity of the  previous event within the distance $r$ from the epicenter of the previous event. We consider events  of their  magnitude  $\mu \geq \mu _c$, and compute the  distribution of the local recurrence time $T$ with $r=7.5$.  The magnitude threshold is set to $\mu_c=2$. 

 In the main panel of Fig. 2, we show on a log-log plot the  computed local recurrence-time distribution function $P(T)$ for the  cases of $\alpha=1$ (a) and of $\alpha=3$ (b). The recurrence time is normalized by its mean $\bar T$, which is $\bar T\nu=2.22$, 0.56, 2.30, 2.67, 2.59 and 3.20 
 (respectively
 for $a=1$, 1/4, 1/8, 1/12, 1/16 and 1/24)  for $\alpha=1$ (Fig. 2(a)), and
 $\bar T\nu=41.4$, 1.72, 2.76, 3.90, 4.81, 6.09 and 6.72 
 (respectively for $a=1$, 1/4, 
 1/8, 1/12, 1/16, 1/24 and 1/32) for $\alpha=3$ (Fig. 2(b)).

 In the case of $\alpha=1$, the recurrence-time distribution $P(T)$ for smaller $a$ exhibits a weak peak at $T/\bar T \sim 0.1$ and a shoulder at $T/\bar{T} \simeq 1$, with a tail of the  distribution showing an exponential behavior at longer $T$. In the case of $\alpha=3$, a smaller-$T$ peak in $P(T)$ becomes more pronounced, while a larger-$T$ shoulder gets almost lost. While the computed $P(T)$ at larger $T$ does not much depend on the grid spacing $a$, the smaller-$T$ peak continues to shift to smaller $T$ as $a$ decreases. We note that, in both cases of $\alpha=1$ and 3, the basic feature of the distribution is robust against the change of the range parameter $r$. 


\subsection{Time correlations of events associated with the mainshock}

 In Fig. 3, we show the time correlation function between large events (mainshock) and events of arbitrary sizes for the cases of $\alpha=1$ (a) and of $\alpha=3$ (b).  In these figures, we plot the mean number of events of arbitrary sizes, dominated in number by small events, occurring within the distance $r=7.5$ from the epicenter of the mainshock before  ($t<0$) and after ($t>0$) the mainshock, where the occurrence of the mainshock is taken to be the origin of the time $t=0$. The average is taken over all large events of their magnitudes of $\mu \geq \mu_c=2$. The number of events are counted here with the time bin of $\Delta t\nu =0.02$.

 In both cases of $\alpha=1$ and 3, qualitative features of the time correlation of the original discrete model, including a prominent seismic acceleration before the mainshock [\textit{Shaw et al.}, 1992; Mori and Kawamura, 2006], persists in the continuum limit. Seismic suppression after the mainshock is more pronounced in $\alpha=3$ than in $\alpha=1$. The data of $\alpha=3$ do not seem to converge to an asymptotic form even at our smallest grid spacing $a=1/32$.

\subsection{Spatial correlations of events before the mainshock}

 In this subsection, we examine the continuum limit of the spatial seismic correlations {\it before\/} the mainshock. Our previous studies on the discrete BK model  [\textit{Mori and Kawamura},  2005; 2006; 2008a] has revealed the occurrence of the doughnut-like quiescence phenomenon prior to the mainshock, {\it i.e.\/}, though the frequency of small events are generally enhanced preceding the mainshock at and around the  epicenter of the upcoming mainshock, the frequency of smaller events is suppressed just before the mainshock in a close vicinity of the upcoming mainshock, while it continues to be enhanced in the surrounding blocks. This phenomenon closely resembles the ``Mogi doughnut'' [\textit{Mogi}, 1969; 1979, \textit{Scholz}, 2002]. The spatial range where the quiescence was observed was narrow, only of a few  blocks. The cause of this doughnut-like quiescence was ascribed to the one-block events  [\textit{Mori and Kawamura}, 2006; 2008a]. 

 Then, a natural question to be addressed is whether the doughnut-like quiescence observed in the discrete BK model  survives the continuum limit, or it is a phenomenon intrinsically originated from the short cut-off length scale of the model. 

 In order to answer this question, we calculate the spatial seismic correlation functions before the mainshock, with systematically varying the grid spacing $a$. The seismic correlation between the mainshock of  $\mu\geq \mu_c=2$ and the preceding events of arbitrary size, dominated in number by small events, are calculated for several time periods  before the mainshock. Fig. 4 exhibits the time-dependent spatial correlation functions in the case of $\alpha=1$, for a larger grid spacing $a=1/4$ (a) and for a smaller grid spacing $a=1/32$ (b), while Fig. 5 exhibits the same quantities in the case of $\alpha=3$. In order to see the dependence on the grid spacing $a$ more directly, we show in Fig. 6 the spatial seismic correlation functions in a shorter time period of $t\nu =0 \sim 0.01$ for various $a$, in the cases of $\alpha=1$ (a) and of $\alpha=3$ (b). As is clear from Fig. 6, the spatial range of the quiesence gets narrower for smaller grid spacing $a$. Namely, while the doughnut-like quiescence  is detectable for smaller grid spacings of $a<1$, as the grid spacing $a$ gets significantly smaller,  however, the spatial range of the  quiescence gets narrower, tending to vanish for small enough $a$. See insets of Fig. 6 where the peak position of the event frequency representing the quiesence scale is plotted as a function of the grid spacing $a$. Apparently, the peak position tends to zero in the continuum limit $a\rightarrow 0$. This observation strongly suggests that the doughnut-like quiescence might vanish altogether in the continuum limit $a \to 0$.

\subsection{The time-dependent magnitude distribution before the mainshock}

 In the original discrete BK model, a significant change of the magnitude distribution is often observed preceding the mainshock [\textit{Mori and Kawamura}, 2005; 2006; 2008a]. In the 1D short-range BK model, in particular, a significant suppression of large events was observed  just before the mainshock, leading to an apparent increase of the effective $B$-value. In this subsection, we examine the continuum limit of such temporal evolution of the magnitude distribution before the mainshock.

 In Figs. 7 and 8, we show the ``time-resolved'' local magnitude distributions for several time periods before the mainshock in the cases of $\alpha=1$ (Fig. 7) and of $\alpha=3$ (Fig. 8).  Figs. (a) and (b) represent the cases of a larger grid spacing $a=1/4$ and a smaller grid spacing $a=1/16$, respectively.  Only events with their epicenters lying within the distance $r=7.5$  from the upcoming mainshock of $\mu\geq \mu_c=2$ are counted here. As can be seen from these figures,  in both cases of $\alpha=1$ and 3, the continuum limit seems to affect only  slightly the temporal evolution of the magnitude distribution. Even in the continuum limit, the suppression of large events is observed before the mainshock, leading to an apparent increase of the effective $B$-value.

\subsection{The time-dependent magnitude distribution after the mainshock}

 Our previous study showed that, in contrast to the behavior before the mainshock, the magnitude distribution of the discrete 1D BK model  {\it after\/} the mainshock showed very little temporal evolution [\textit{Mori and Kawamura}, 2006; 2008a]. Here, we examine how the continuum limit affects this property after the mainshock.

 In Figs. 9 and 10, we show the time-resolved local magnitude  distributions after the mainshock for the case $\alpha=1$ (Fig. 9) and of $\alpha=3$ (Fig. 10).  Figs. (a) and (b) represent the cases of a larger grid spacing $a=1/4$ and a smaller grid spacing $a=1/16$, respectively. Other conditions are the same as those in Figs. 7 and 8.  As can be seen from these figures, in both cases of $\alpha=1$ and 3, the magnitude distribution  in the continuum limit shows little-to-no temporal evolution after the mainshock.  Aftershock sequence obeying the Omori law is never realized in the model even in its continuum limit.

\section{The viscosity model}

 In this section, we show the results of our numerical simulations on the 1D BK model with a nonzero viscosity term $\eta=0.02$ for various observables studied in the previous section.  As in the previous section, the velocity-weakening parameter $\alpha$ is taken to be either $\alpha=1$ or $\alpha=3$, whereas the parameter $\sigma$ is fixed to $\sigma=0.01$.

\subsection{The magnitude distribution}

 The computed magnitude distributions $R(\mu)$ are shown in Figs. 11(a) and (b) for the cases of  $\alpha=1$ and 3, respectively. As can be seen from the figures, a nonzero viscosity tends to weaken the GR character of the magnitude distribution. Namely, it weakens the linearity of the $R(\mu)$ curve, causing an eminent down-bending behavior. The main cause of this deviation from the GR law comes from the supression of events at smaller magnitudes in the viscousity model relative to those in the non-viscousity model. Indeed, the viscosity tends to make the relative displacement of neighboring blocks  being smoothe, enhancing the correlated motion of neighboring blocks. As a result, the frequency of smaller events of one or a few blocks is considerably reduced in the viscous model, which causes the observed deviation from the GR law at smaller magnitudes.

 Notable difference from the non-viscous case shown in Fig. 1 is that $R(\mu)$ seems to converge to an asymptotic form for smaller $a$ in both cases of $\alpha=1$ and 3, except that the minimum magnitude continuously gets lower as the grid spacing $a$ gets smaller. A similar result was already reported by [\textit{Shaw}, 1994].

 The small-length cut-off scale $\epsilon$ as given by Eq.(9) is estimated here to be $\epsilon \simeq 0.44$ and 0.26 for $\alpha=1$ and 3, respectively. As can be seen from Figs. 11(a) and (b), $R(\mu)$ converges to an asymptotic form for the $\alpha$-values smaller than $a \simeq 1/4$ and 1/8 for $\alpha=1$ and 3, respectively, which is consistent with the expected condition of the continuum limit $a < \epsilon$.

 In  Fig. 12, the viscosity $\eta$ dependence of $R(\mu)$ is shown with fixing $a=1/4$.  In both cases of $\alpha=1$ and 3, as the viscosity gets larger, weights of larger events are significantly suppressed. In the case of $\alpha=1$ shown in Fig. 12(a), $R(\mu)$ exhibits a down-bending curvature, while in the case of $\alpha=3$ shown in Fig. 12(b), the peak of $R(\mu)$ at a larger magnitude shifts toward a smaller magnitude.

\subsection{The local recurrence-time distribution}

 In Fig. 13, we show on a log-log plot the  local recurrence-time distribution function $P(T)$ for the cases of $\alpha=1$ (a) and $\alpha=3$ (b). Here, we consider large events of their magnitude  $\mu \geq \mu _c=2$, and compute  the local  recurrence time $T$ with the range parameter $r=7.5$. The recurrence time is normalized by its mean  $\bar T$, which is
$\bar T\nu=2.22$, 1.85, 1.88, 1.82, 1.82, 1.85 and 1.85 
 (respectively for $a=1$, 1/4, 1/8, 1/12, 1/16, 1/24 and 1/32) 
 for $\alpha=1$ (Fig. 13(a)), and $\bar T\nu=39.8$, 1.44, 2.01, 2.03, 2.04, 2.06
 and 2.03 (respectively for
 $a=1$, 1/4, 1/8, 1/12, 1/16, 1/24 and 1/32) for $\alpha=3$ (Fig. 13(b)).

 The behavior of the recurrence-time distribution $P(T)$ in the viscous case is qualitatively similar to that in the non-viscous case, although the convergence is much faster here in the viscous case. In the case of $\alpha=1$, the asymptotic $P(T)$ exhibits a weak peak at $T/\bar T \sim 0.1$ and a  shoulder at $T/\bar{T} \simeq 1$, with a tail of the  distribution showing an exponential behavior at longer $T$.  In the case of $\alpha =3$, as can be seen from Fig. 13(b), the asymptotic $P(T)$ exhibits a peak at $T\simeq 0.2$.


\subsection{Time correlations of events associated with the mainshock}

 In Fig. 14, we show the time correlation function between
large events (mainshock) and events of arbitrary sizes, dominated in
number by small events, for the cases of $\alpha=1$ (a) and $\alpha=3$ (b). 
The other conditions are the same as in Fig. 3 in the non-viscous case.

 Again, with decreasing the grid spacing $a$, the results converge to the continuum limit fairly quickly. The obtained behavior of the time correlation looks more or less similar to the one in the non-viscous case, although the convergence is much faster here in the viscous case even including the case of $\alpha=3$.

\subsection{Spatial correlations of events before the mainshock}

 In this subsection, we examine the spatial seismic
correlations {\it before\/} the mainshock. The calculational conditions are the same as in Figs. 4-6 in the viscous case. The computed spatial seismic correlation function  immediately before the mainshock in the shorter time period of $t\nu=0\sim0.01$ is shown in Fig. 15 for  various grid spacings $a$. The frictional parameter is either $\alpha=1$ (a) or $\alpha=3$ (b). 

 The doughnut-like quiescence  is appreciable even for smaller grid spacings of $a<1$ for both cases of $\alpha=1$ and 3. As in the non-viscous case studied in the previous section,  as the grid spacing $a$ gets smaller,  the spatial range of the  quiescence gets narrower, tending to vanish for small enough $a$: See the insets of Fig. 15. This observation strongly suggests again that the doughnut-like quiescence might vanish altogether in the continuum limit $a \to 0$.

 Hence, the doughnut-like quiescence observed in the discrete BK model  is likely to be a phenomenon closely related to the short-length cut-off scale of the model. This seems fully consistent with our observation in the previous papers that the one-block events are responsible  for the observed doughnut-like quiescence [\textit{Mori and Kawamura}, 2006; 2008a].

\subsection{The time-dependent magnitude distribution before the mainshock}

 In Figs. 16 and 17, we show the ``time-resolved'' local magnitude
 distributions for several time periods
 before the mainshock. The case of 
 $\alpha=1$ is shown in Fig. 16, and the case of
 $\alpha=3$ in Fig. 17. 
 Figs. (a) and (b) represent the cases of a larger grid spacing $a=1/4$ and a smaller grid spacing $a=1/16$, respectively. The calculational conditions are the same as in Figs. 7 and 8 of the non-viscous case.

In both cases of $\alpha=1$ and 3, as can be seen from Figs. 16 and 17, the temporal evolution of $R(\mu)$ prior to the mainshock is kept qualitatively similar even when  the grid spacing $a$ gets smaller. Namely, on approaching the mainshock, weights of larger events are significantly suppressed. Similar behavior was observed  in Figs. 7 and 8 for the non-viscous case.

 We have also studied the ``time-resolved'' local magnitude  distributions after the mainshock (the results not shown here). As in the non-viscous case shown in Figs. 9 and 10, the magnitude distribution hardly changes on approaching the mainshock even for smaller grid spacing $a$.

\bigskip

 Overall, the statistical properties of the viscosity BK model in the continuum limit turn out to be qualitatively similar to those of the non-viscosity BK model in the continuum limit, although the ``critical'' aspect, {\it e.g.\/}, the GR-like behavior, is weakened somewhat. By contrast, the convergence to the asymptotic continuum limit is much faster in the viscous case than in the non-viscous case. The manner of this convergence is well controlled by the length scale given by Eq.(3).

\section{Summary and discussion}

 Spatiotemporal correlations of the one-dimensional spring-block (Burridge-Knopoff) model of earthquakes, either with or without the viscosity term, are studied by means of numerical computer simulations. The continuum limit of the model is examined by systematically investigating the model properties with varying the block-size parameter $a$ toward $a\rightarrow 0$. 

 The two types of models are analyzed, {\it i.e.\/}, the 1D BK model with the Kelvin viscosity and the one without the Kelvin viscosity. The naive continuum limit of the BK model without the viscosity is known to be problematic in that the dynamics becomes pathological. The Kelvin viscosity term is introduced so that the model dynamics possesses a sensible continuum limit. The added viscosity term introduces a small-length cut-off scale $\epsilon = \pi \sqrt{\frac{\eta^{\prime}}{\alpha \omega}}$, which indeed turns out to control the continuum limit of the model. It also turns out that the viscosity affects the statistical properties of the model somewhat, making large earthquakes smaller and enhancing the deviation from the GR law. In other words, the viscosity suppresses the critical feature of the model. The viscosity tends to make the relative displacement of neighboring blocks being smoother and enhance the correlated motion of blocks.

 Probably reflecting the intrinsic problem associated with its pathological dynamics, even the {\it statistical properties\/} of the non-viscosity BK model sometimes continues to change even at our smallest grid spacing studied. In contrast, those of the viscosity BK model are found to converge well to an asymptotic continuum limit once the grid spacing is taken smaller than the above small-length cut-off scale $\epsilon$. One obvious consequence of the continuum limit $a\rightarrow 0$ limit is that the size of minimum earthquakes gets smaller. It thus appears that, in the continuum limit of the model, even an infinitesimal earthquake is possible. By contrast, many of the statistical properties of the original discrete BK model are kept qualitatively the same even in the continuum limit: For example, the model exhibits a seismic acceleration preceding the mainshock. The magnitude distribution of the model exhibits a significant temporal change before the mainshock, while it exhibits only a negligible change after the mainshock. This observation in turn gives some guarantee that the original discrete BK model provides a reasonable description of the continuum fault.

 One notable difference between the original discrete model and the corresponding model in the continuum limit is the existence/non-existence of the doughnut-like quiescence just before the mainshock.  Large events of the original discrete BK model exhibits a seismic acceleration before the mainshock, which is accompanied  by a doughnut-like quiescence occurring just before the mainshock in close vicinity of the upcoming mainshock. It is found that in both viscous and non-viscous cases, as the grid spacing $a$ gets smaller, the spatial  range of the doughnut-like quiescence becomes narrower, and the doughnut-like quiescence  might vanish altogether in the continuum limit. The doughnut-like quiescence observed in the  discrete BK model is then a phenomenon closely related to the short cut-off length scale.

 This observation might have some implications to real
 seismicity. While the real crust is  obviously a continuum, it is not
 so obvious whether there exists an  intrinsic short-length cut-off
 scale in real seismicity or not. In any case, in real earthquakes,
 the ``Mogi-doughnut'' is occasionally reported to occur [\textit{Mogi}, 1969; 1979, \textit{Scholz}, 2002], although to  elucidate its statistical significance is often not easy. Our present result might suggest that, if the real crust possesses a cut-off length scale, the  ``Mogi- doughnut'' quiescence might occur at such a length scale. In other words, spatial inhomogeneity might be an essential ingredient for the ``Mogi-doughnut'' to occur.

 For the future, it is desirable to extend the present analysis of the continuum limit of the 1D BK model to the corresponding 2D model. Since taking the $a\rightarrow 0$ limit necessarily means taking the large-size limit,  performing such analysis in 2D is computationally demanding. It is also desirable to perform a similar analysis for the BK model based on the rate- and state-dependent friction law [\textit{Morimoto and Kawamura}, 2007], which is the friction law  employed in the study of Rice [\textit{Rice}, 1993]. This friction law naturally provides a length scale into the problem, even without invoking the Kelvin viscosity. Thus, in order to fully resolve the question raised by Rice, it would be necessary to perform a similar analysis for the BK model with the rate- and state-dependent friction law.  We leave such interesting extension of the present analysis to a future task.

\setfigurenum{1}
\begin{figure}[ht]
\begin{center}
\includegraphics[scale=0.65]{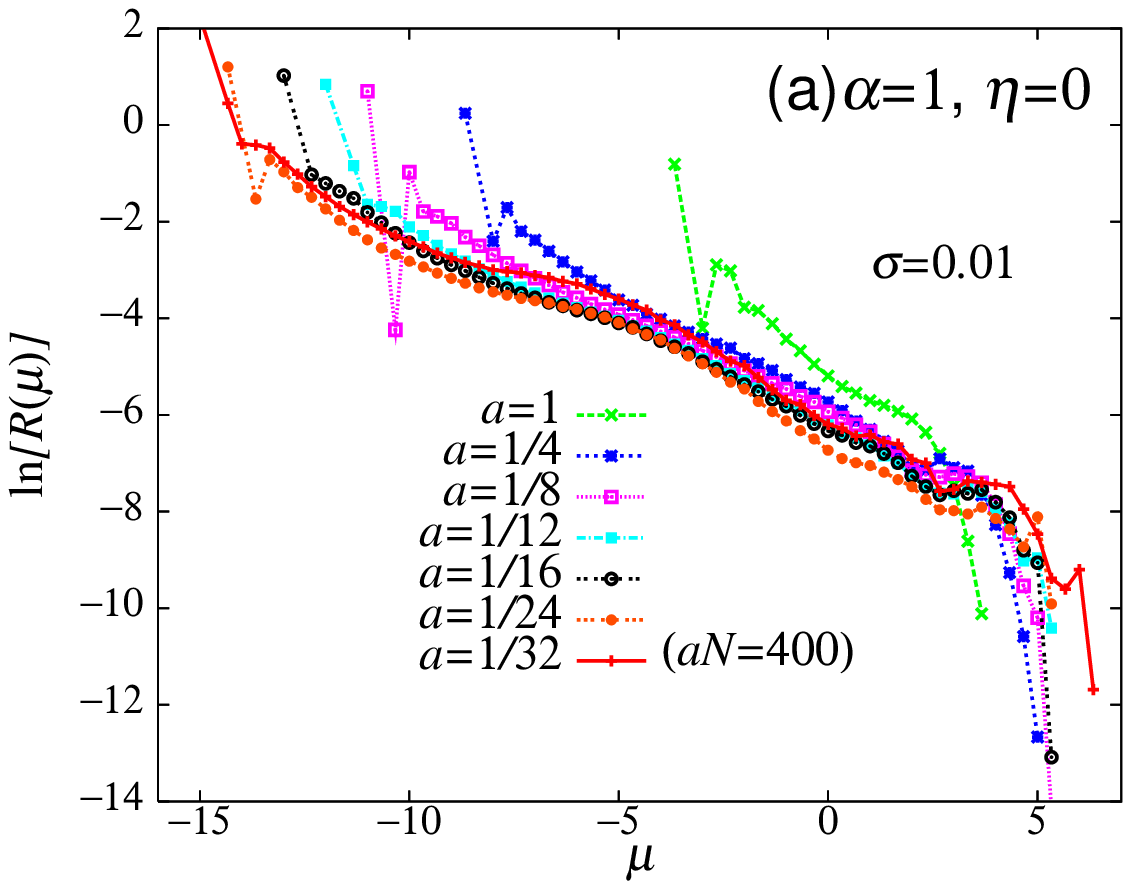}
\includegraphics[scale=0.65]{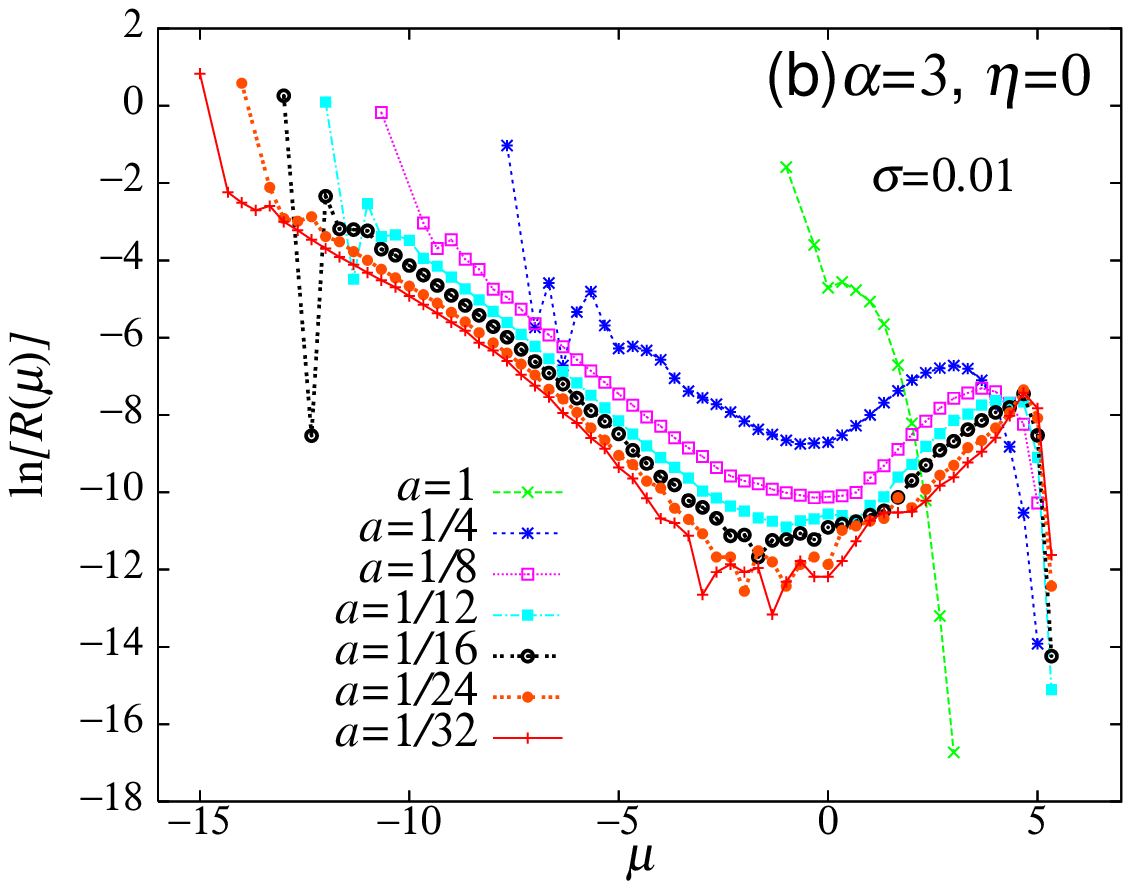}
\end{center}
\caption{
The magnitude distribution $R(\mu)$ of earthquake events of the  1D non-viscosity BK model ($\eta=0$). The dimensionless grid spacing  $a$ is varied in the range $1 \geq a \geq 1/32$. Figs. (a) and (b) represent the cases of $\alpha=1$ and 3, respectively.  The parameter $\sigma$ is fixed to $\sigma=0.01$.  The system size is $L=aN=200$, except the case of $a=1/32$ of $\alpha=1$ where $L$ is taken to be $L=aN=400$ to circumvent the more severe finite-size effect.
}
\end{figure}

\newpage

\setfigurenum{2}
\begin{figure}[ht]
\begin{center}
\includegraphics[scale=0.65]{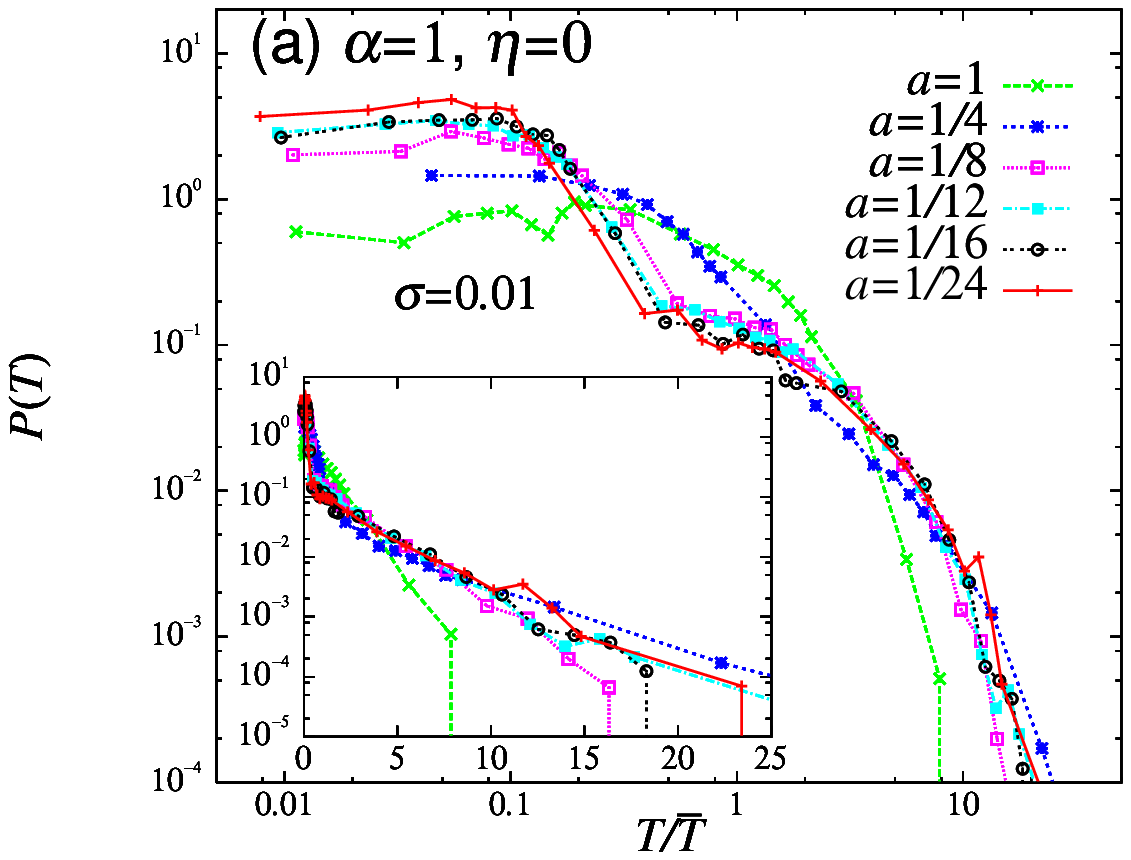}
\includegraphics[scale=0.65]{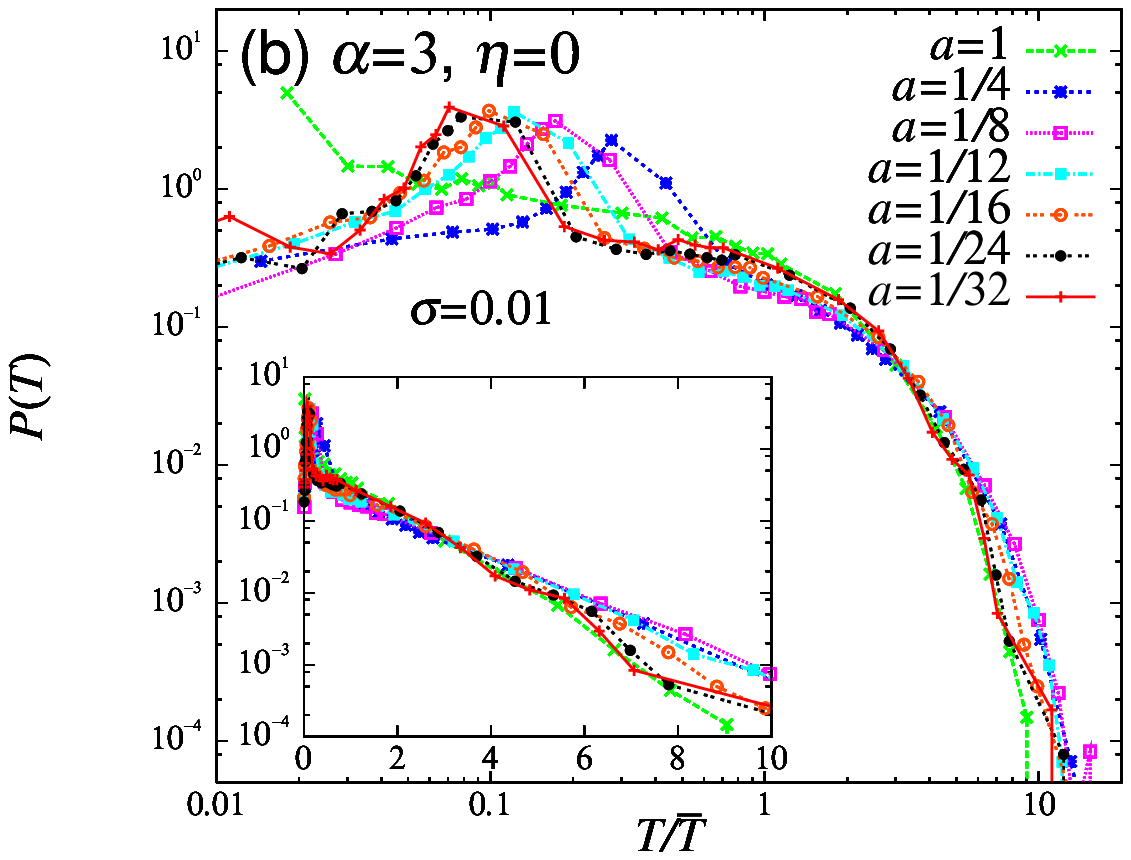}
\end{center}
\caption{
 The local recurrence-time distribution function $P(T)$ of the 1D
 non-viscosity BK model ($\eta=0$), with  varying the dimensionless
 grid spacing $a$.  Figs. (a) and (b) represent  the cases of
 $\alpha=1$ and 3, respectively.  The parameter $\sigma$ is fixed to
 $\sigma =0.01$.  The main panels represent the log-log plots of
 $P(T)$, while  the insets represent the semi-logarithmic plots
 including the  tail part of the distribution. The mean recurrence
 time $\bar T$ is  $\bar T\nu=2.22$, 0.56, 2.30, 2.67, 2.59 and 3.20 
 (respectively
 for $a=1$, 1/4, 1/8, 1/12, 1/16 and 1/24)  for $\alpha=1$, and
 $\bar T\nu=41.4$, 1.72, 2.76, 3.90, 4.81, 6.09 and 6.72 
 (respectively for $a=1$, 1/4, 
 1/8, 1/12, 1/16, 1/24 and 1/32) for $\alpha=3$.
}
\end{figure}

\newpage

\setfigurenum{3}
\begin{figure}[ht]
\begin{center}
\includegraphics[scale=0.65]{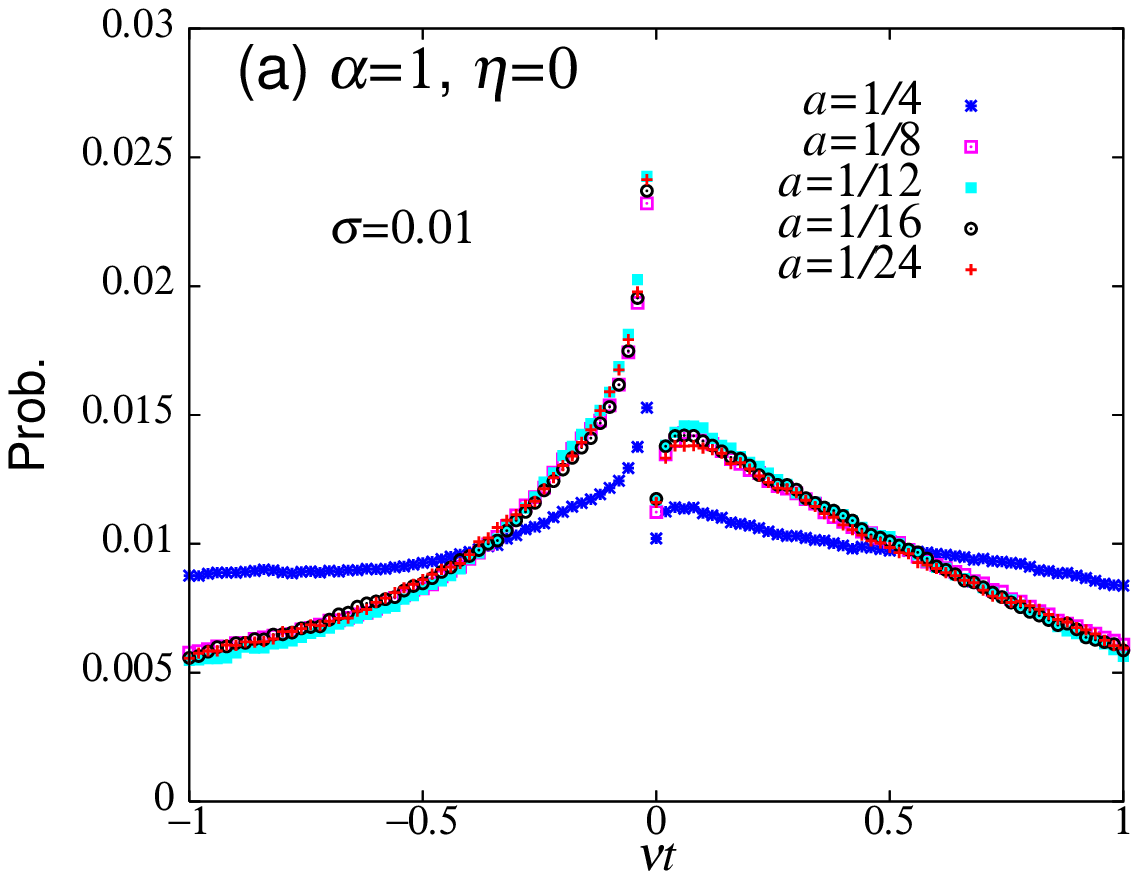}
\includegraphics[scale=0.65]{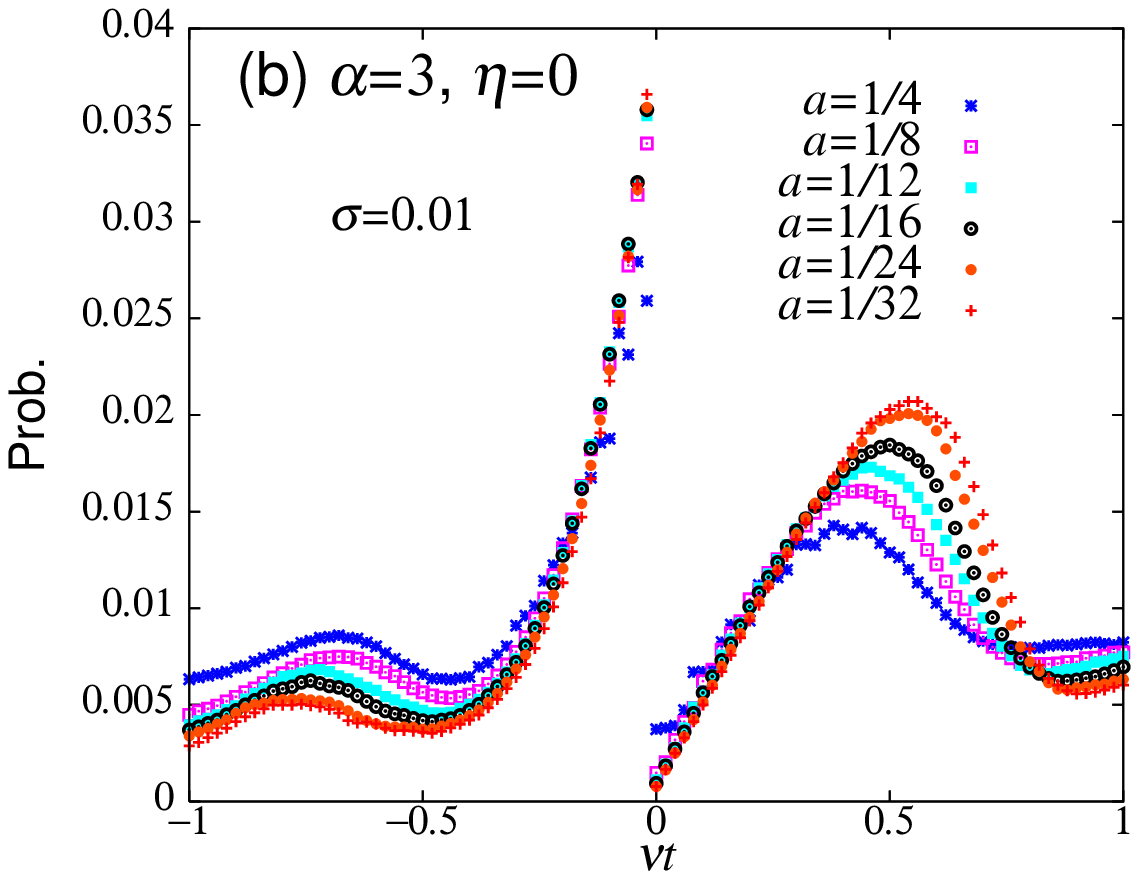}
\end{center}
\caption{
The time correlation function of the 1D non-viscosity BK model ($\eta=0$)  
 between large events of $\mu_c=2$ (mainshock) occurring at
 time $t=0$ and
events of arbitrary sizes (dominated in number by small events) occurring at
time $t$. The dimensionless grid spacing $a$ is varied in the range  $1/4 \geq a \geq 1/32$. Fig. (a) represents the case of $\alpha=1$, while
Fig. (b) represents the case of $\alpha=3$. 
The parameter $\sigma$ is fixed to $\sigma =0.01$.
Events of arbitrary sizes occurring within the distance 
$r= 7.5$ from the epicenter of the mainshock are counted. 
The negative time $t<0$ represents the time before the mainshock, 
while the positive time $t>0$ represents the time after the mainshock. 
The average is taken over all large events of its magnitude 
$\mu >\mu_c=2$. The system size is $L=aN=200$.
}
\end{figure}

\newpage

\setfigurenum{4}
\begin{figure}[ht]
\begin{center}
\includegraphics[scale=0.65]{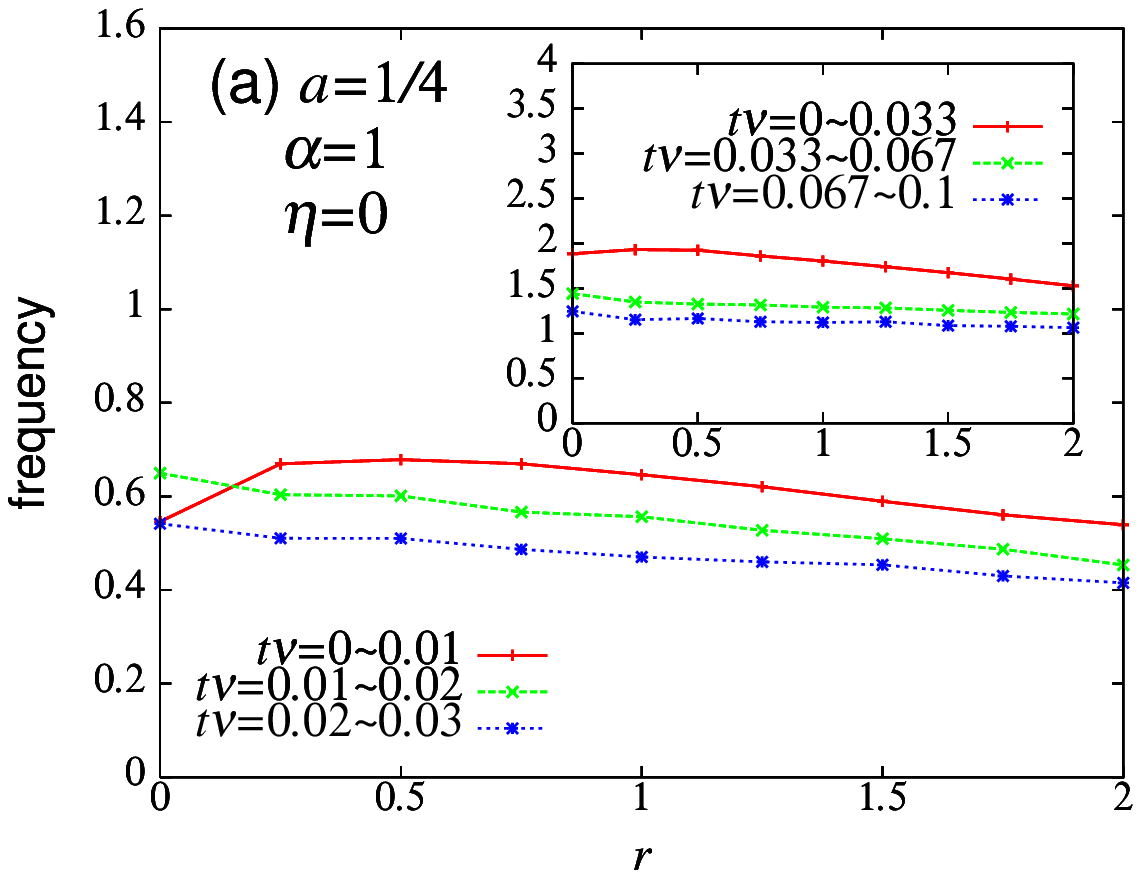}
\includegraphics[scale=0.65]{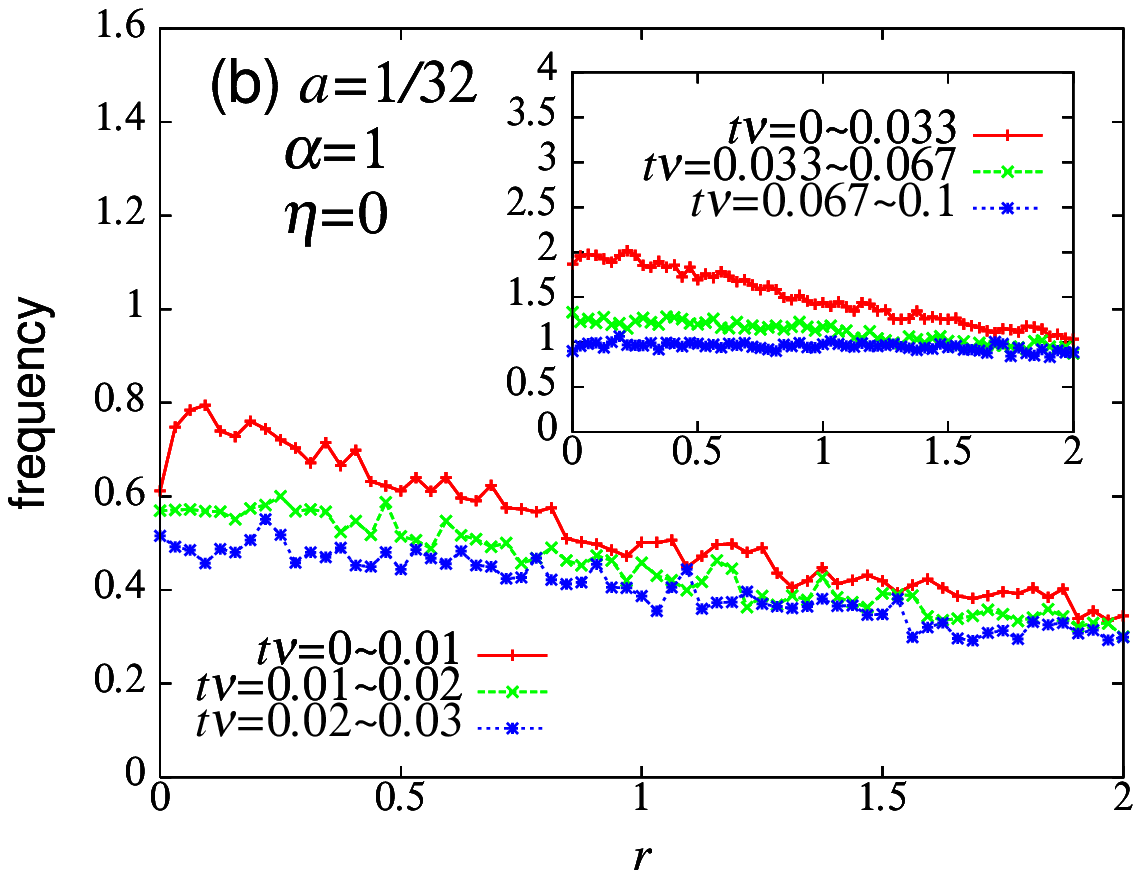}
\end{center}
\caption{
Event frequency for several  time periods before the mainshock of $\mu >\mu_c=2$ of the 1D non-viscosity BK model  ($\eta=0$) plotted versus $r$, the distance from the epicenter of the upcoming mainshock. The frictional parameters are $\alpha=1$ and $\sigma =0.01$. The dimensionless grid spacing $a$ is $a=1/4$ (a), and $a=1/32$ (b). The system size is $L=aN=200$. The insets represent similar plots at longer times.
}
\end{figure}

\newpage

\setfigurenum{5}
\begin{figure}[ht]
\begin{center}
\includegraphics[scale=0.65]{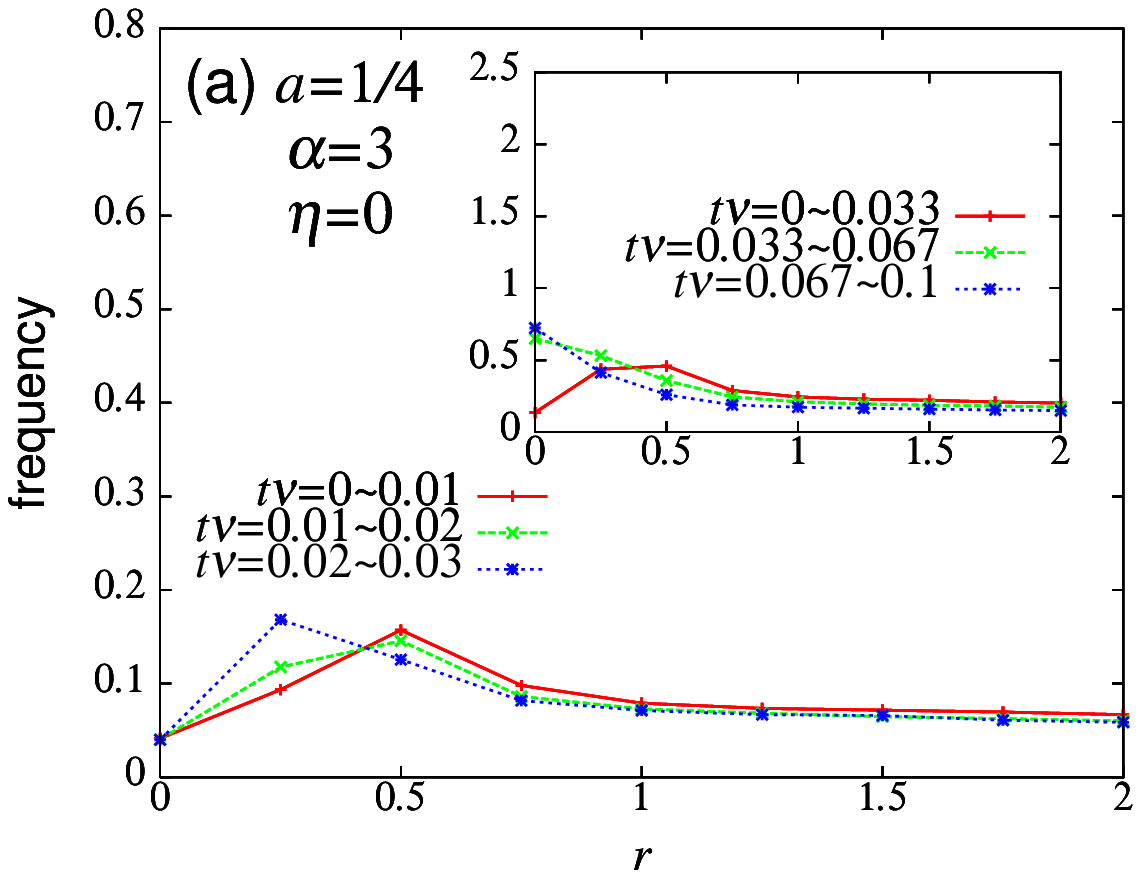}
\includegraphics[scale=0.65]{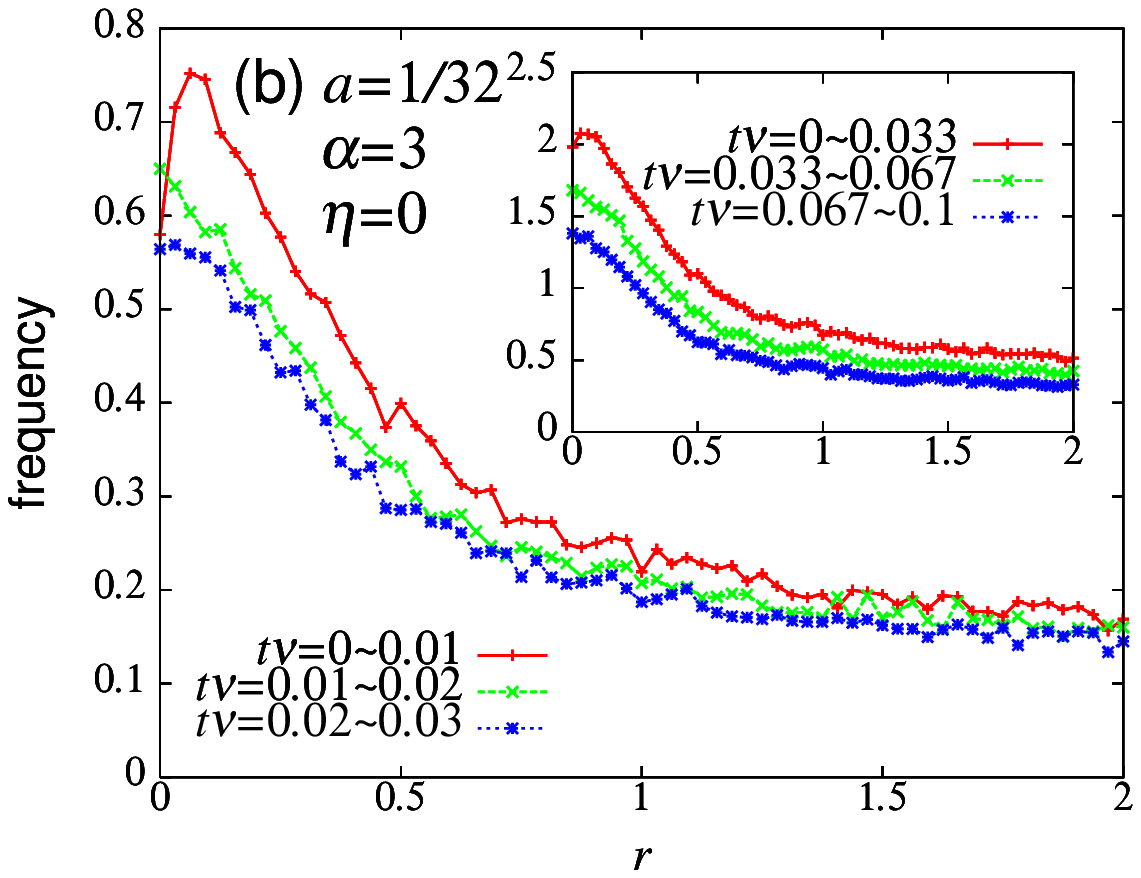}
\end{center}
\caption{
Event frequency for several  time periods before the mainshock of $\mu >\mu_c=2$  of the 1D non-viscosity BK model  ($\eta=0$) plotted versus $r$, the distance from the epicenter of the upcoming mainshock. The frictional parameters are $\alpha=3$ and $\sigma =0.01$. The dimensionless grid spacing $a$ is $a=1/4$ (a), and $a=1/32$ (b). The system size is $L=aN=200$. The insets represent similar plots at longer times.
}
\end{figure}

\newpage

\setfigurenum{6}
\begin{figure}[ht]
\begin{center}
\includegraphics[scale=0.65]{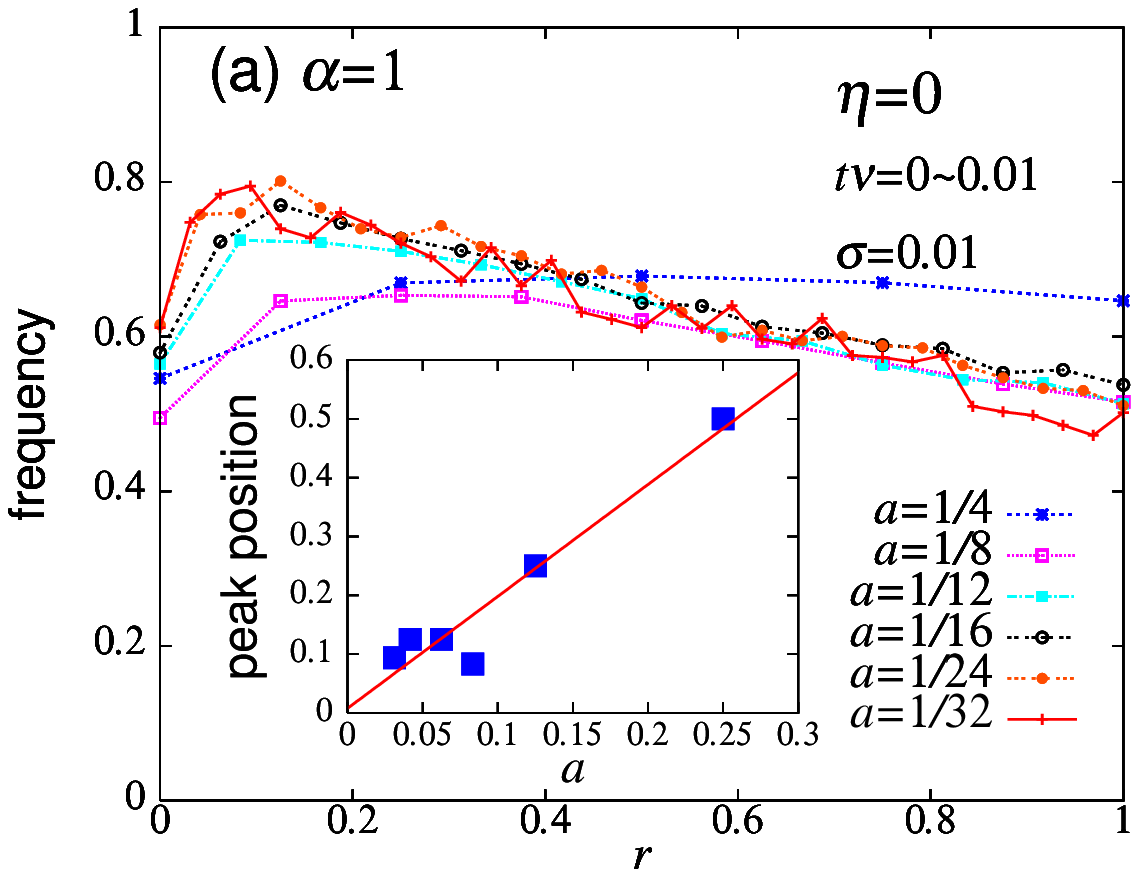}
\includegraphics[scale=0.65]{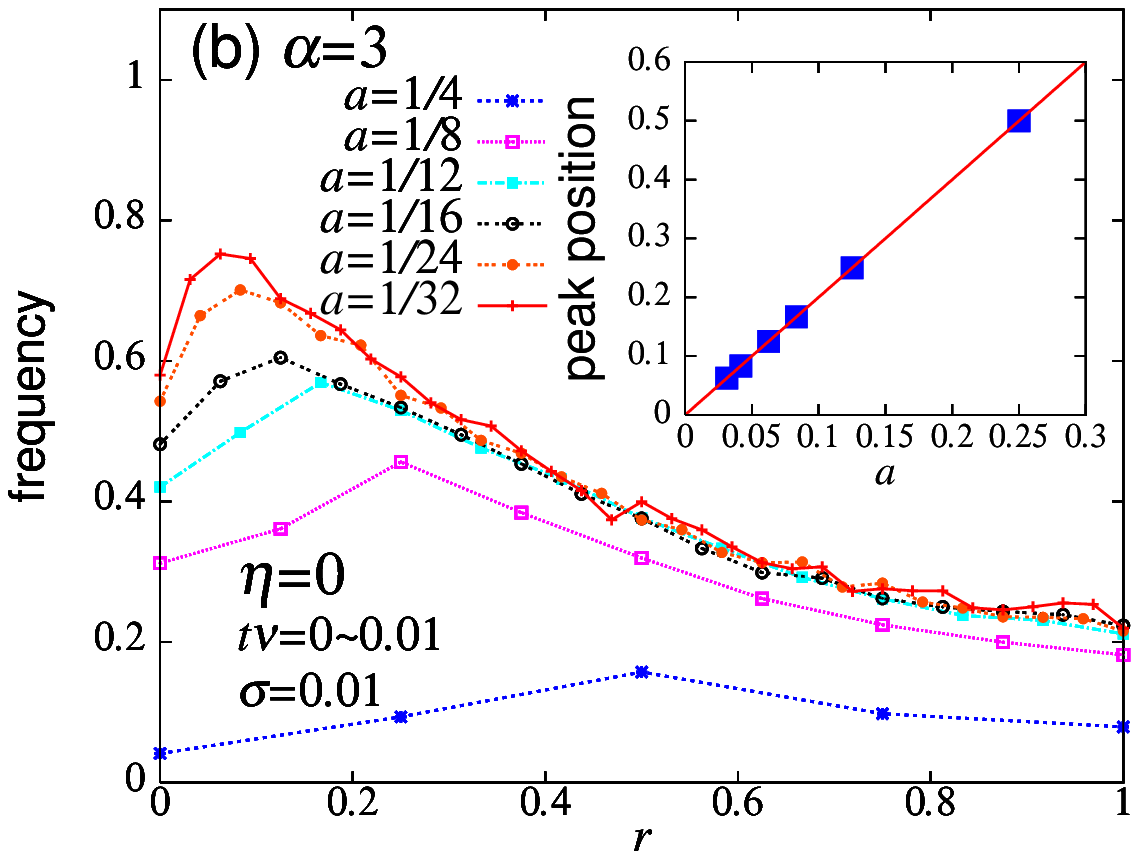}
\end{center}
\caption{
 Event frequency in the time period $t\nu=0\sim0.01$ 
 immediately before the mainshock of $\mu >\mu_c=2$
  of the 1D non-viscosity BK model ($\eta=0$) plotted versus $r$,
 the distance from the epicenter of the upcoming mainshock.  Fig. (a) represents the case of $\alpha=1$, while Fig. (b) represents the case of $\alpha=3$. The dimensionless grid spacing $a$ is  varied in the range $1/32 \leq a \leq 1/4$. The parameter $\sigma$ is
 fixed to $\sigma=0.01$ The system size is $L=aN=200$. The insets represent the peak position of the event frequency, corresponding to the range of the doughnut-like quiescence, as a function of the dimensionless grid spacing $a$. In both cases of $\alpha=1$ and 3, the doughnut-like quiescence appears to vanish in the continuum limit $a\rightarrow 0$.
}
\end{figure}

\newpage

\setfigurenum{7}
\begin{figure}[ht]
\begin{center}
\includegraphics[scale=0.65]{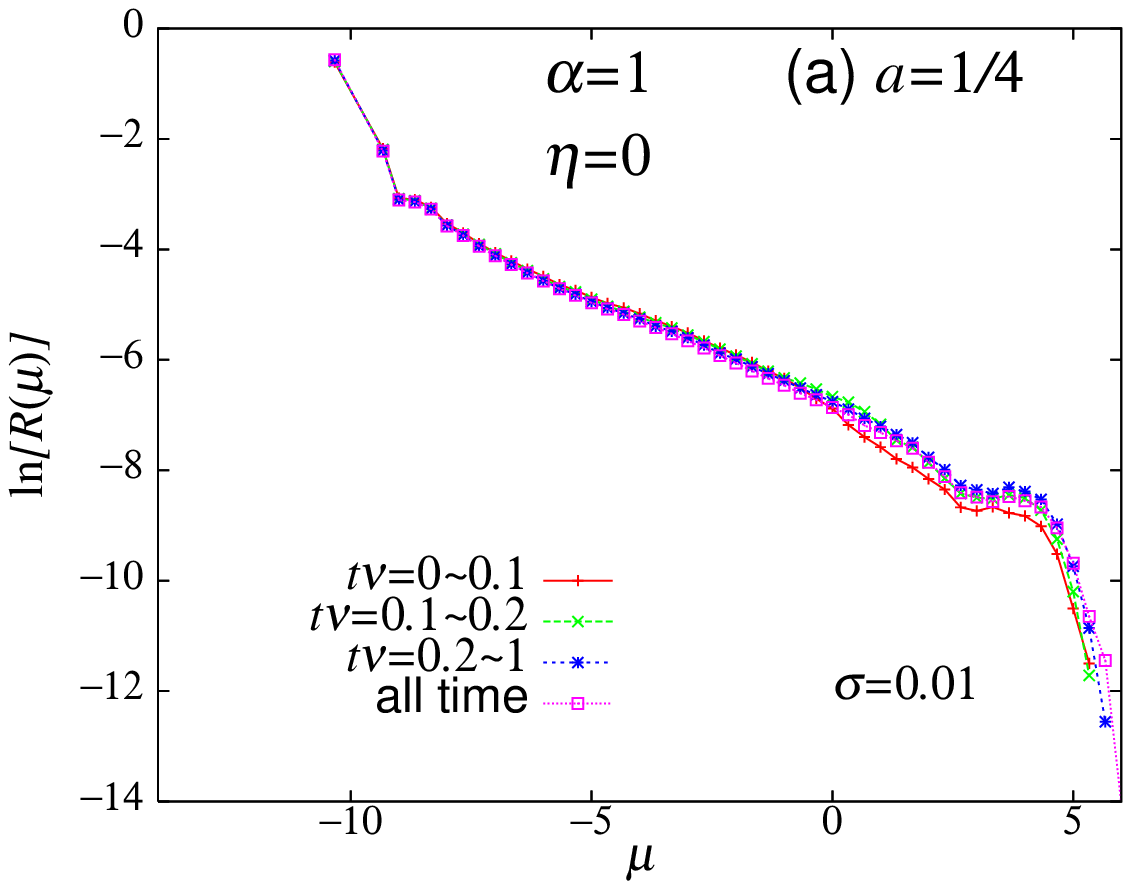}
\includegraphics[scale=0.65]{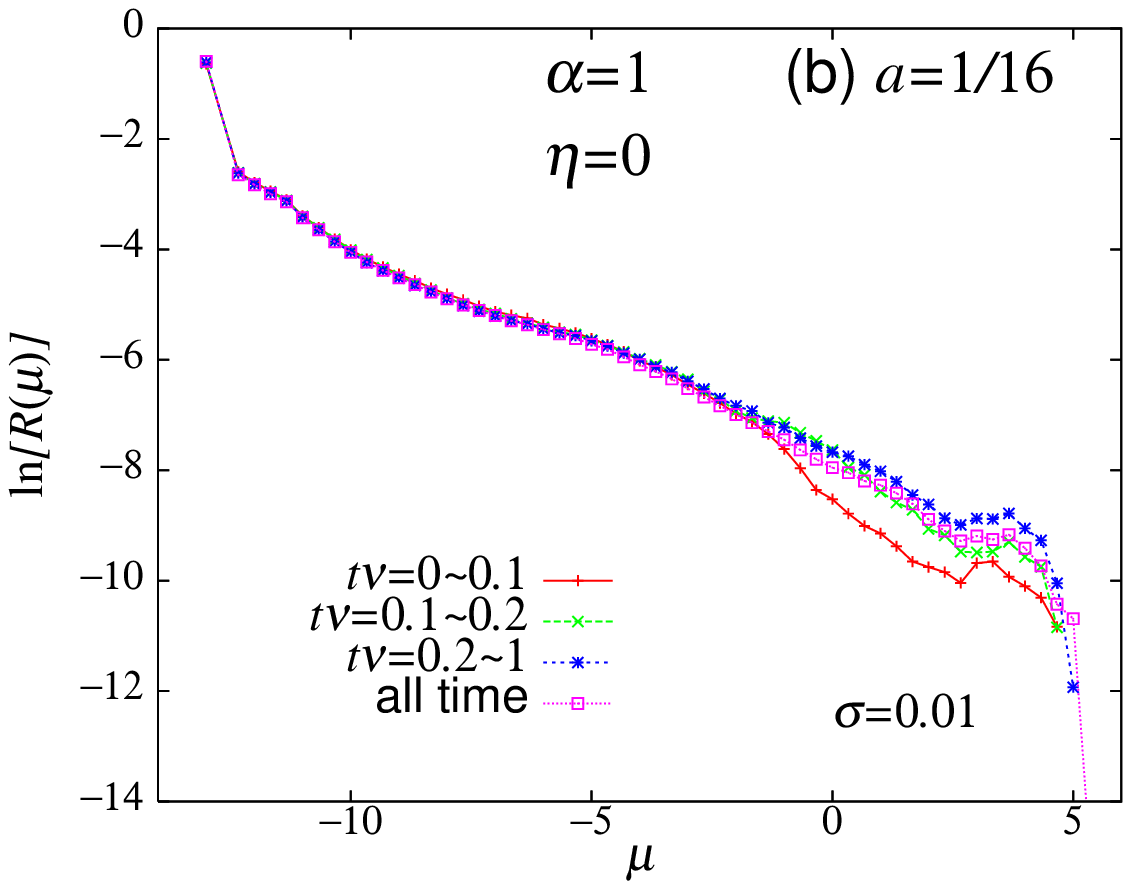}
\end{center}
\caption{
The local magnitude distribution for several time periods before the mainshock  of $\mu >\mu _c=2$ of the 1D non-viscosity BK model
($\eta=0$). The frictional parameters are $\alpha=1$ and  $\sigma=0.01$. Figs. (a) and (b) represent the cases of $a=1/4$ (a)  and $a=1/16$ (b), respectively. Events whose epicenter lies within the distance $r=7.5$ from the epicenter of the upcoming mainshock are counted. The system size is $L=aN=200$.
}
\end{figure}

\newpage

\setfigurenum{8}
\begin{figure}[ht]
\begin{center}
\includegraphics[scale=0.65]{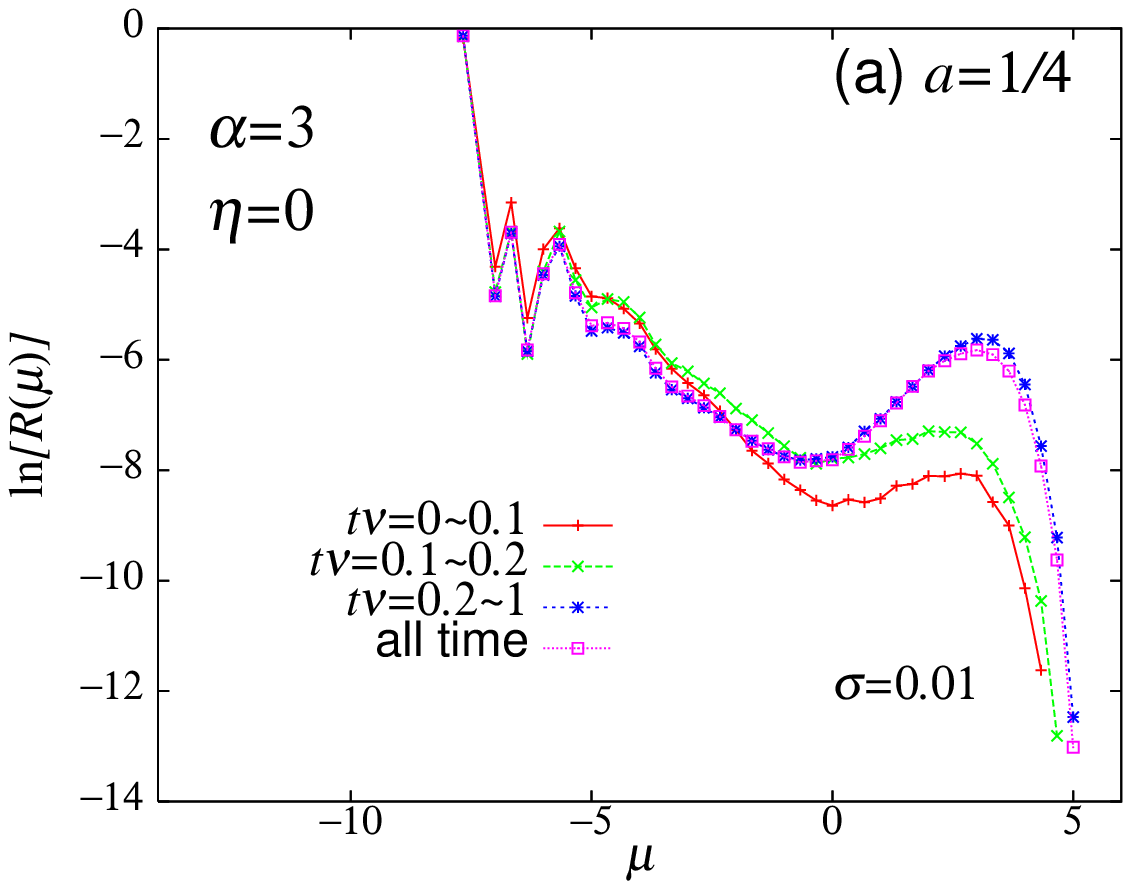}
\includegraphics[scale=0.65]{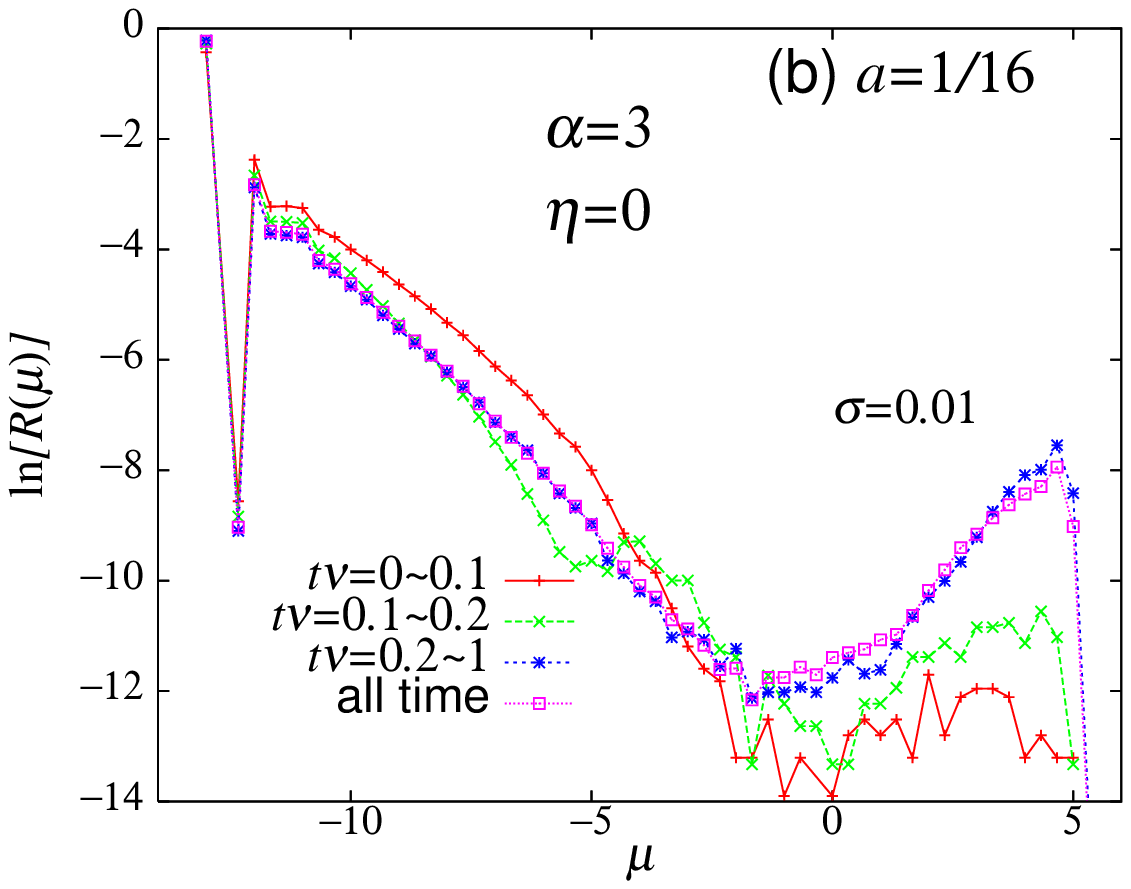}
\end{center}
\caption{
The local magnitude distribution for several time periods before the
mainshock of $\mu >\mu _c=2$ of the 1D non-viscosity BK model 
($\eta=0$).  The frictional parameters are $\alpha=3$ and  $\sigma=0.01$. Figs. (a) and (b) represent the cases of $a=1/4$ (a) and $a=1/16$ (b), respectively. Events whose epicenter lie within the distance $r=7.5$ from the epicenter of the upcoming mainshock are counted. The system size is $L=aN=200$. 
}
\end{figure}

\newpage

\setfigurenum{9}
\begin{figure}[ht]
\begin{center}
\includegraphics[scale=0.65]{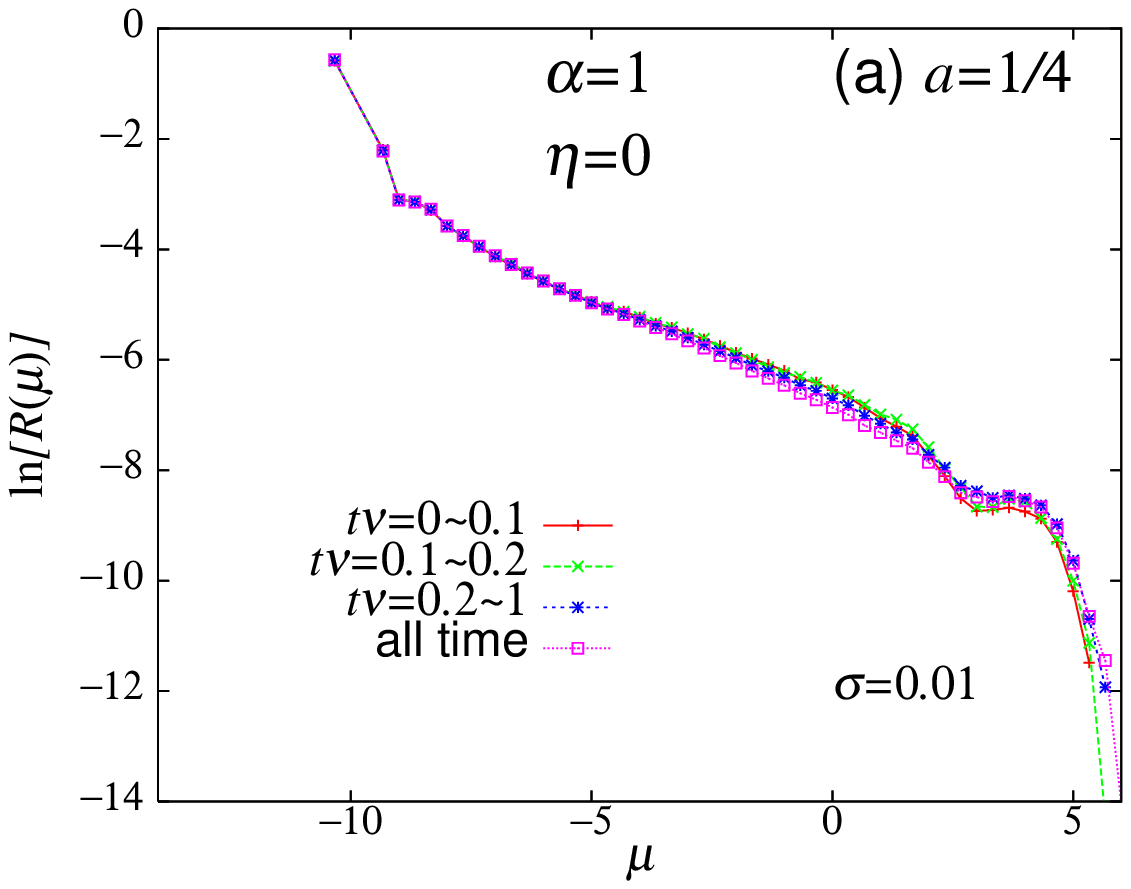}
\includegraphics[scale=0.65]{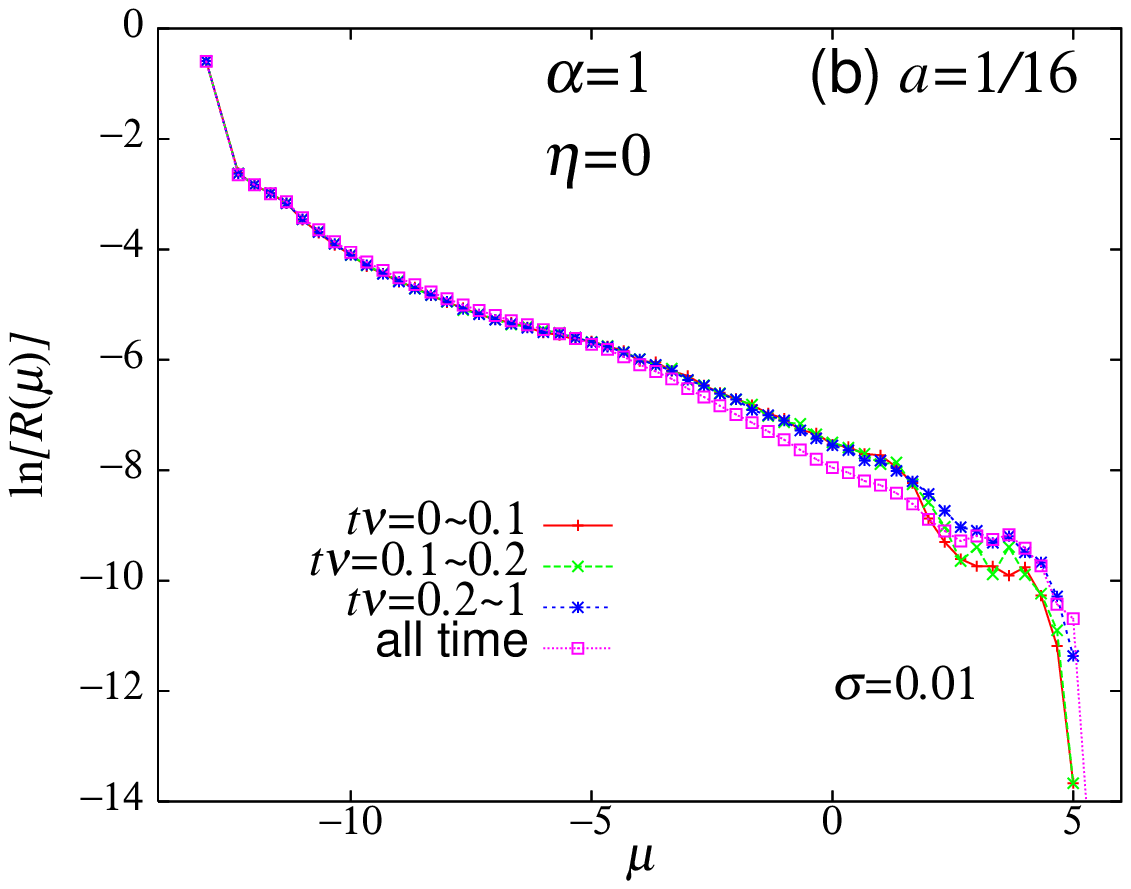}
\end{center}
\caption{
The local magnitude distribution for several time periods after the
mainshock of $\mu >\mu _c=2$ of the 1D non-viscosity BK model
($\eta=0$). The frictional parameters are $\alpha=1$ and  $\sigma=0.01$. Figs. (a) and (b) represent the cases of $a=1/4$ (a) and $a=1/16$ (b), respectively. Events whose epicenter lie within the distance $r=7.5$ from the epicenter of the preceding mainshock are counted. The system size is $L=aN=200$. 
}
\end{figure}

\newpage

\setfigurenum{10}
\begin{figure}[ht]
\begin{center}
\includegraphics[scale=0.65]{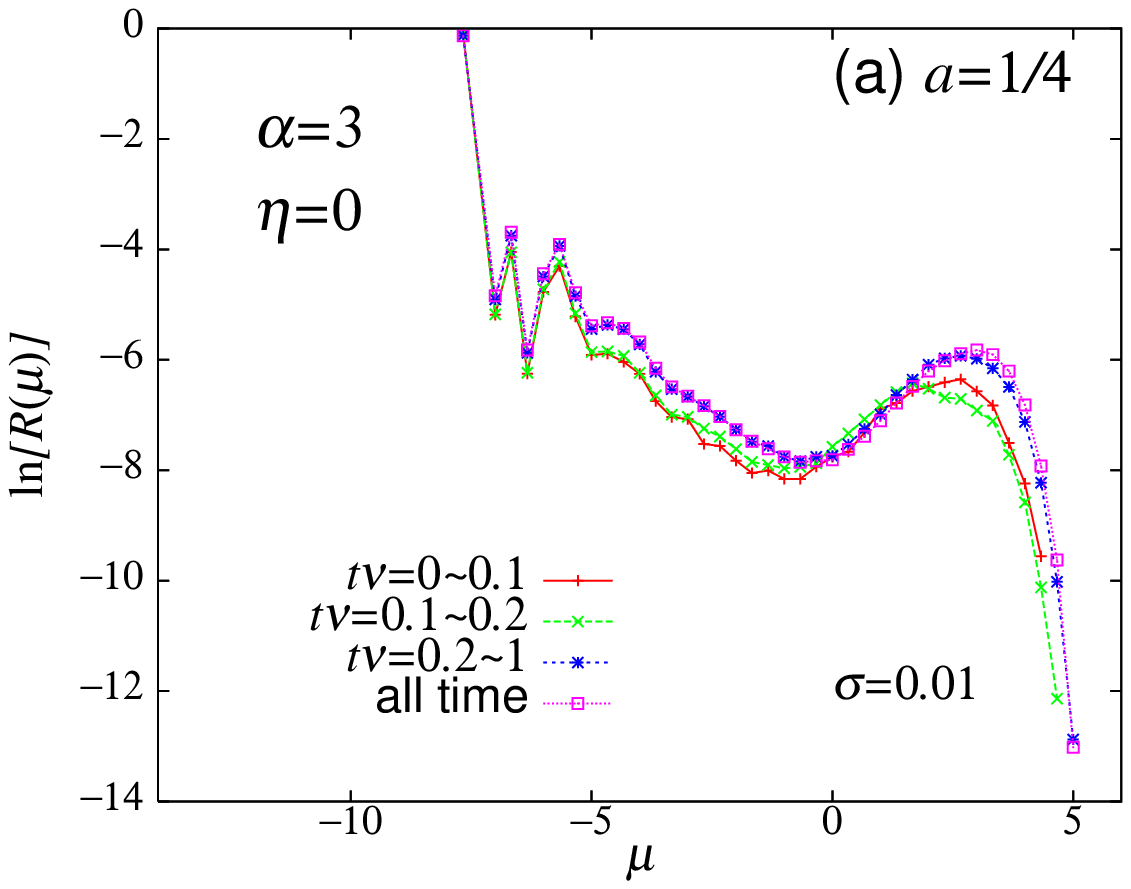}
\includegraphics[scale=0.65]{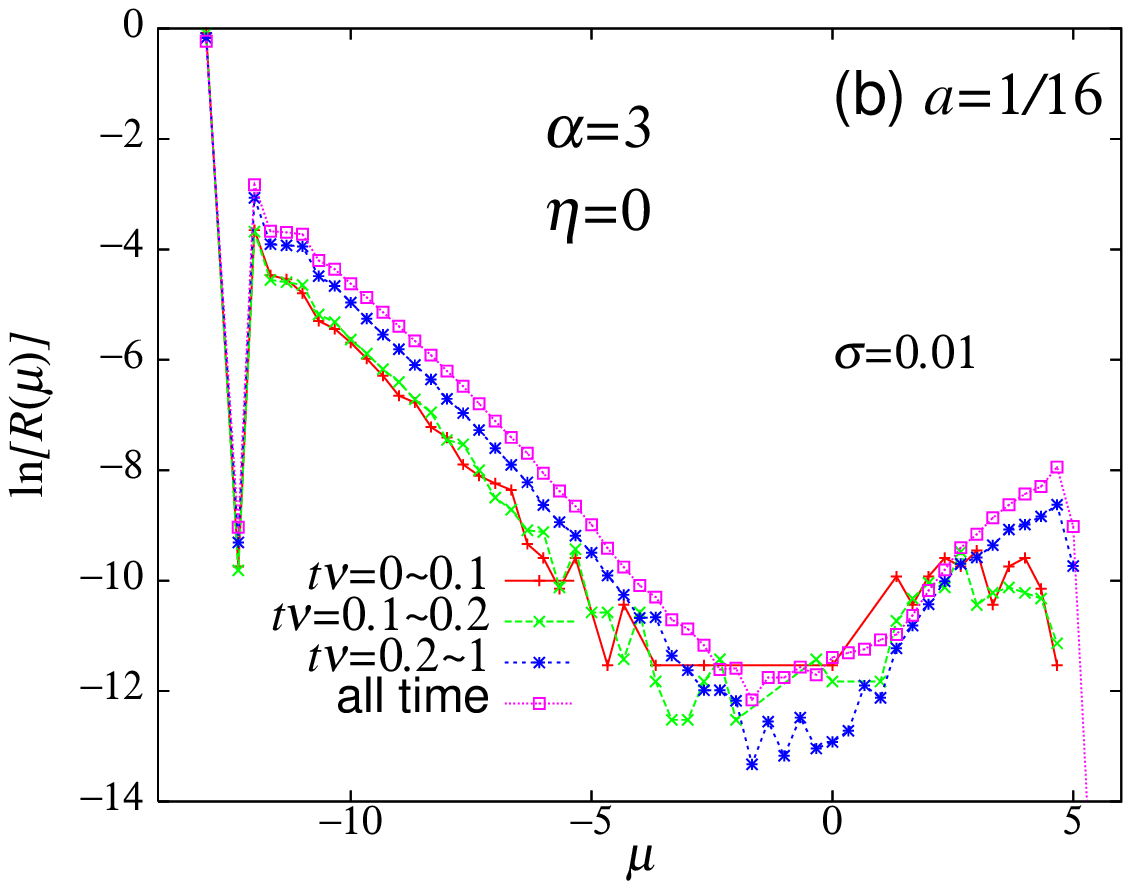}
\end{center}
\caption{
The local magnitude distribution for several time periods after the mainshock of $\mu >\mu _c=2$ of the 1D non-viscosity BK model ($\eta=0$). The frictional parameters are $\alpha=3$ and  $\sigma=0.01$. Figs. (a) and (b) represent the cases of $a=1/4$ (a) and $a=1/16$ (b), respectively. Events whose epicenter lie within the distance $r=7.5$ from the epicenter of the preceding mainshock are counted. The system size is $L=aN=200$. 
}
\end{figure}

\newpage

\setfigurenum{11}
\begin{figure}[ht]
\begin{center}
\includegraphics[scale=0.65]{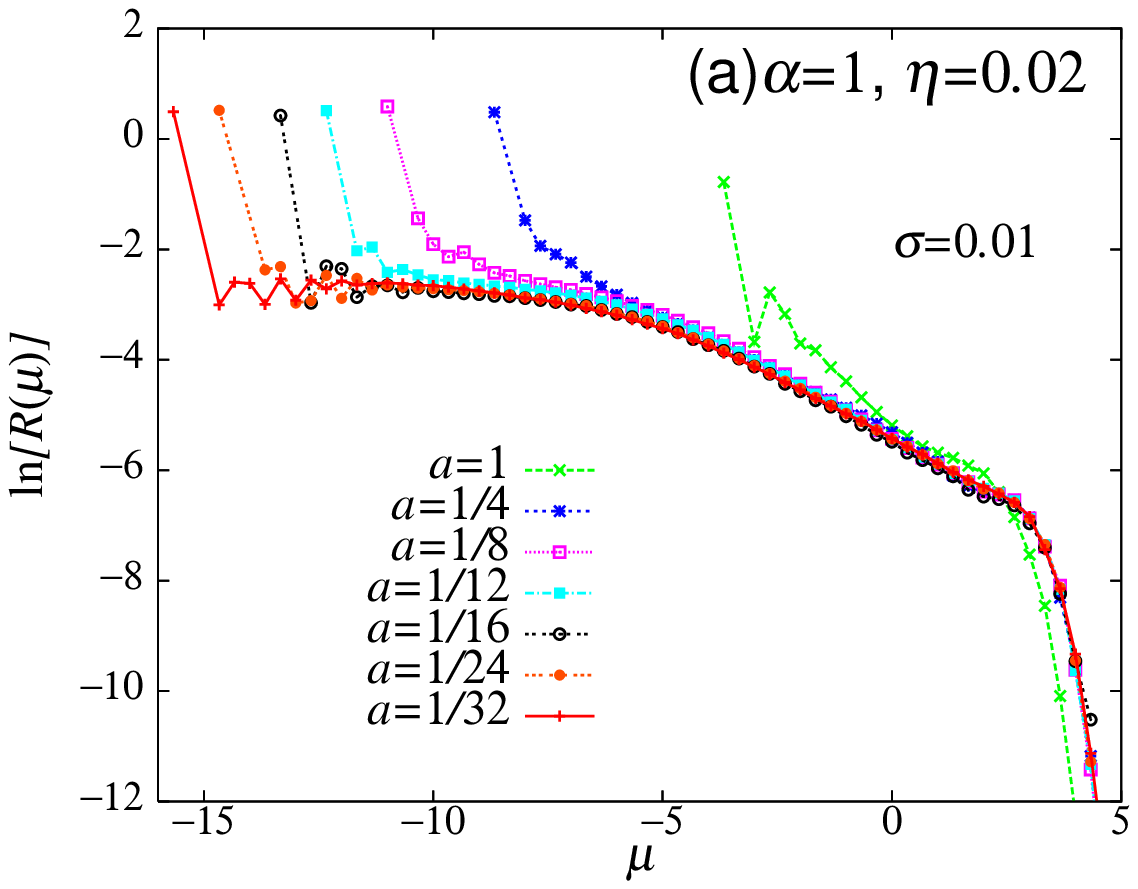}
\includegraphics[scale=0.65]{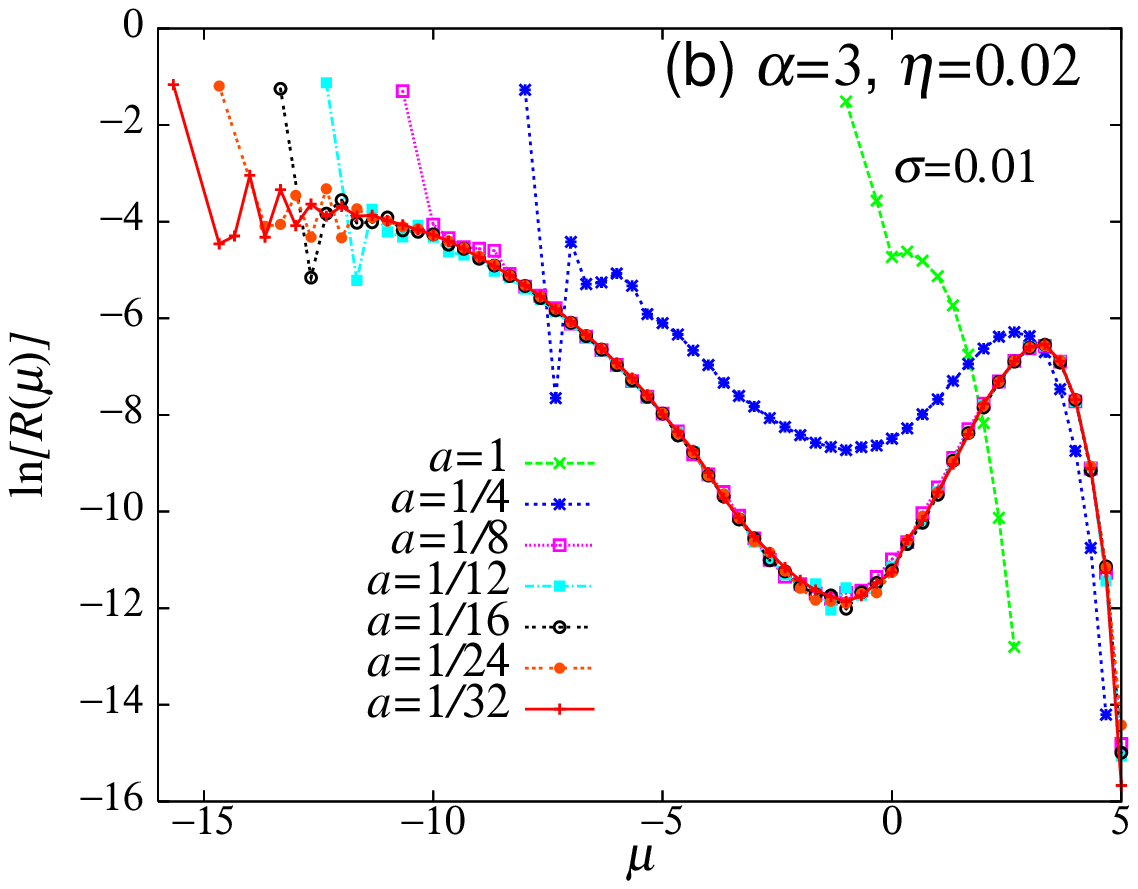}
\end{center}
\caption{
The magnitude distribution $R(\mu)$ of earthquake events of the 
 1D viscosity BK model ($\eta=0.02$) with $\sigma=0.01$. The dimensionless grid spacing  $a$ is varied in the range $1 \geq a \geq 1/32$. Figs. (a) and (b) represent  the cases of $\alpha=1$ and 3, respectively. 
 The system size is $L=aN=200$.
}
\end{figure}

\newpage

\setfigurenum{12}
\begin{figure}[ht]
\begin{center}
\includegraphics[scale=0.65]{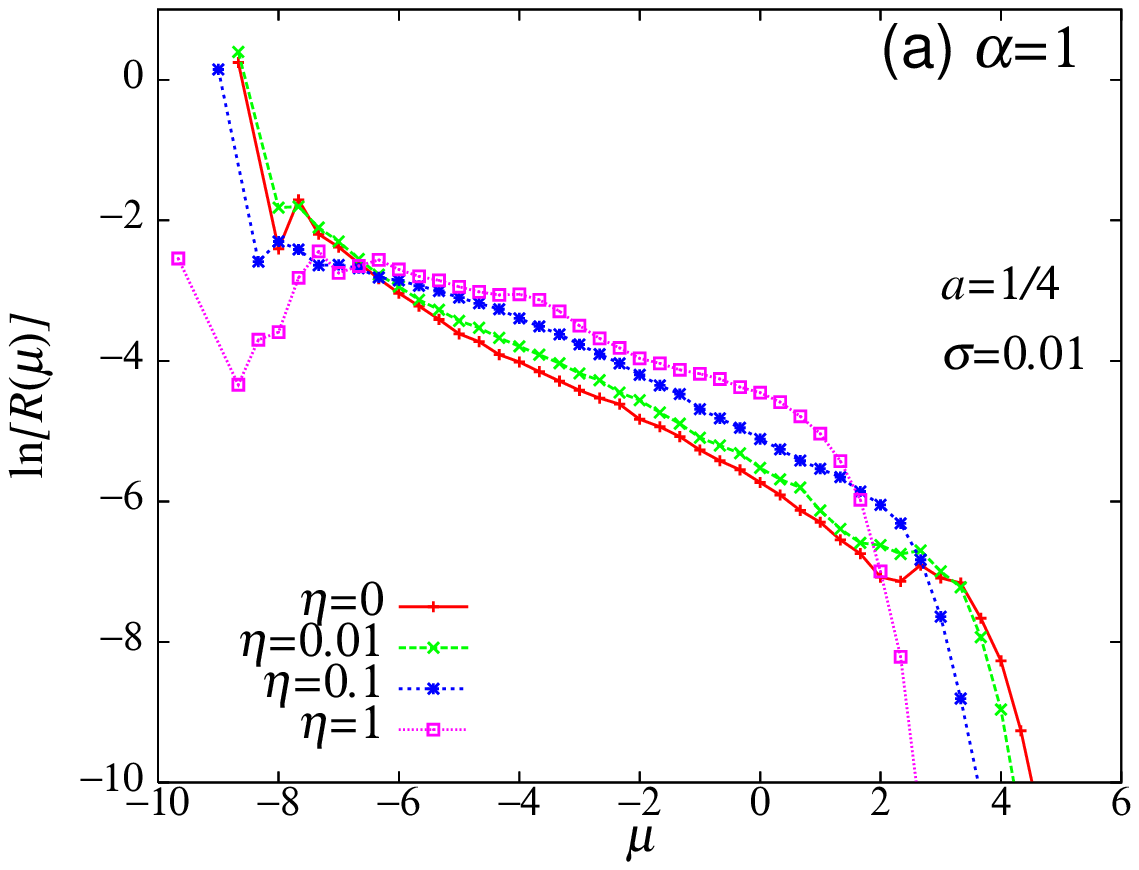}
\includegraphics[scale=0.65]{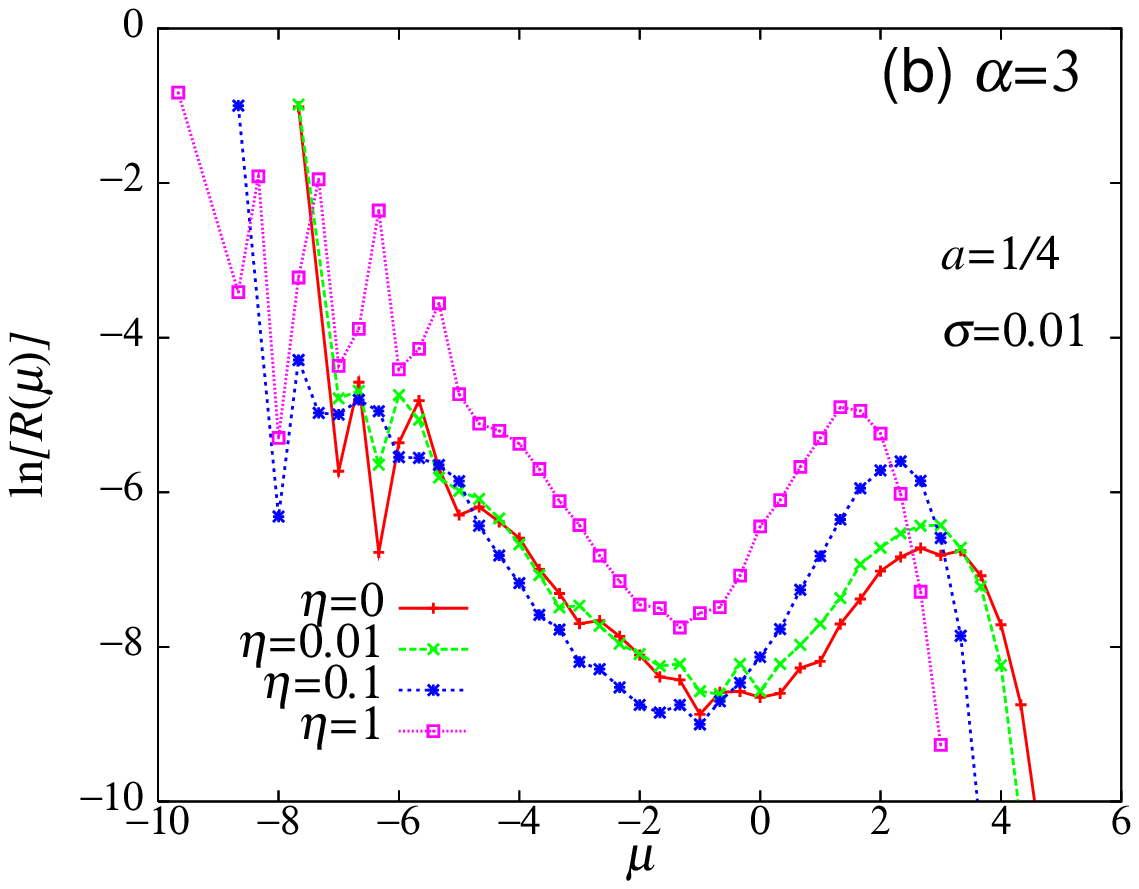}
\end{center}
\caption{
The magnitude distribution $R(\mu)$ of earthquake events of the 
 1D viscosity BK model for various values of the viscosity parameter
 $\eta$. Figs. (a) and (b) represent the cases of $\alpha=1$ and 3,
 respectively. The other parameters are fixed to $a=1/4$ and $\sigma=0.01$. 
 The system size is $L=aN=200$.
}
\end{figure}

\newpage

\setfigurenum{13}
\begin{figure}[ht]
\begin{center}
\includegraphics[scale=0.65]{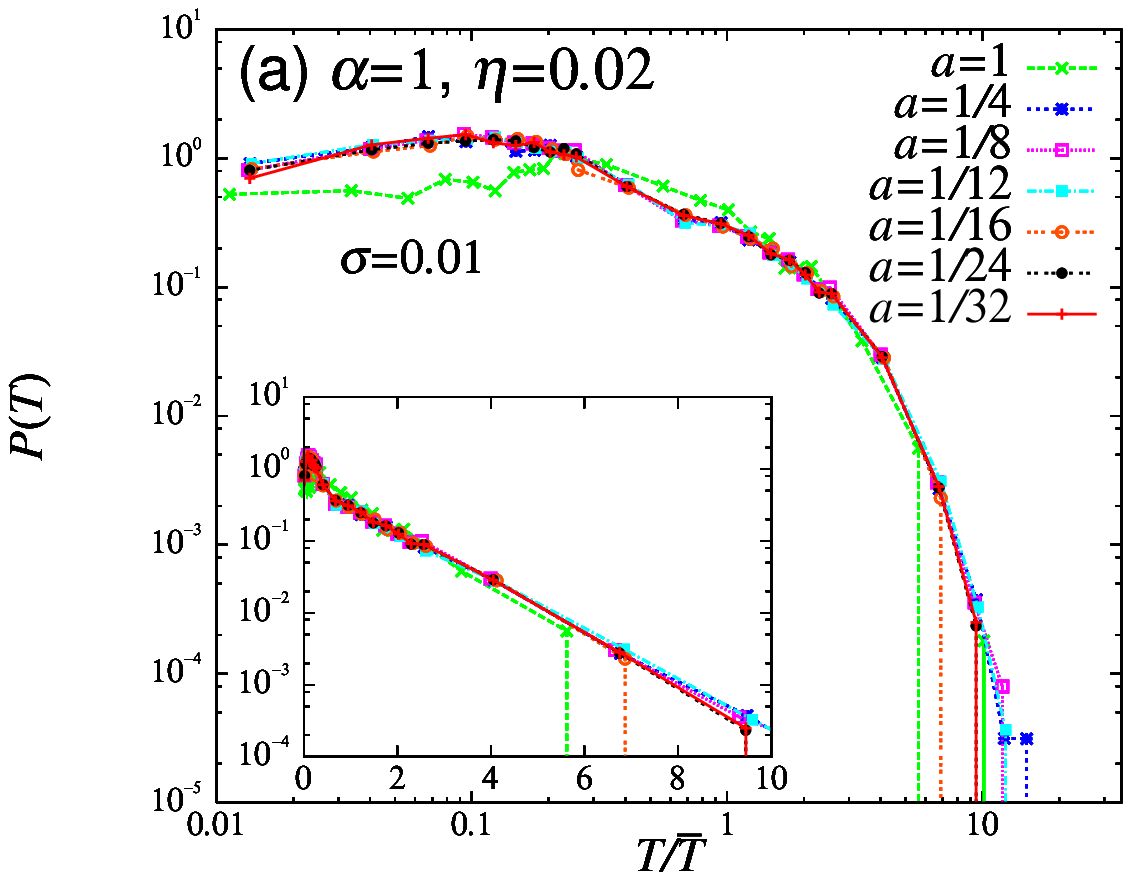}
\includegraphics[scale=0.65]{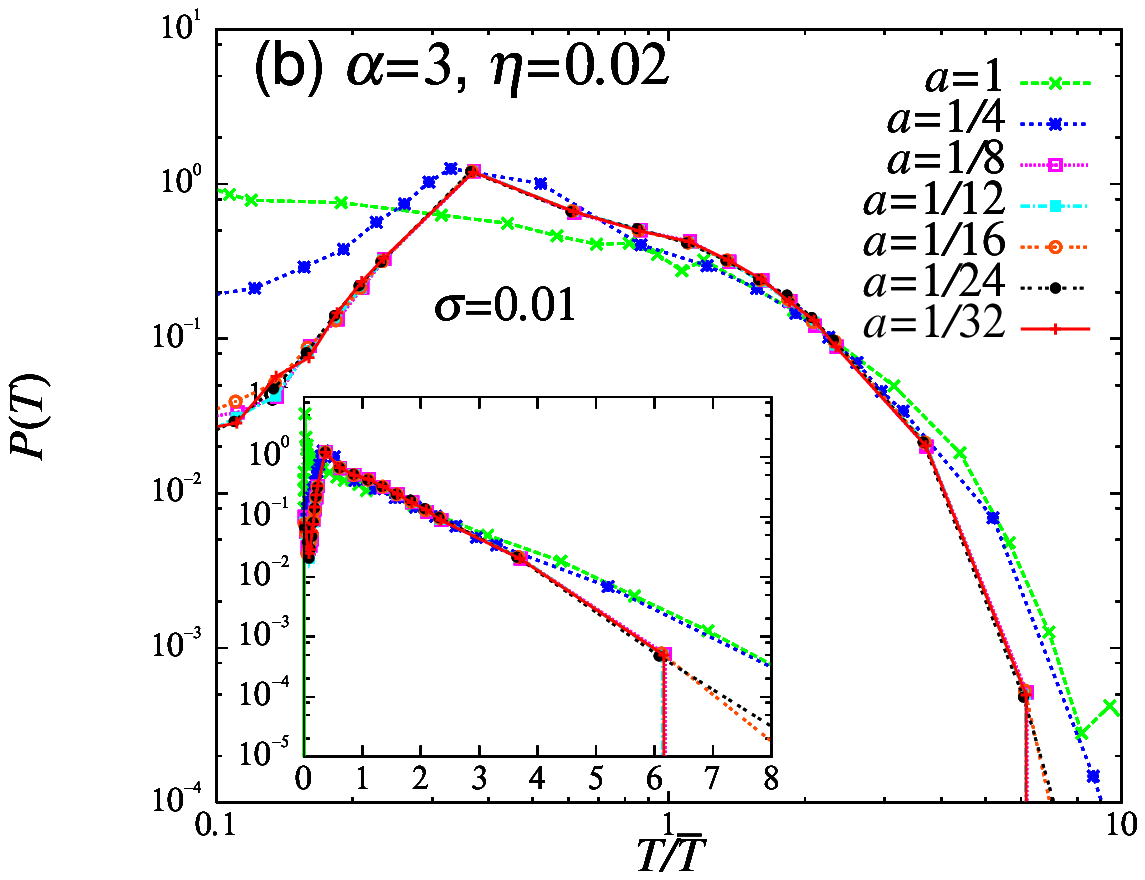}
\end{center}
\caption{
 The local recurrence-time distribution function $P(T)$ of the 1D
 viscosity BK model ($\eta=0.02$), with
 varying the dimensionless grid spacing $a$.  Figs. (a) and (b) represent  the case of $\alpha=1$ and 3, respectively.  The parameter $\sigma$ is fixed to $\sigma =0.01$. The main panels represent the log-log plots of $P(T)$, while 
 the insets represent the semi-logarithmic plots including the 
 tail part of the distribution. The mean recurrence time $\bar T$ is
 $\bar T\nu=2.22$, 1.85, 1.88, 1.82, 1.82, 1.85 and 1.85 
 (respectively for $a=1$, 1/4, 1/8, 1/12, 1/16, 1/24 and 1/32) 
 for $\alpha=1$, and $\bar T\nu=39.8$, 1.44, 2.01, 2.03, 2.04, 2.06
 and 2.03 (respectively for
 $a=1$, 1/4, 1/8, 1/12, 1/16, 1/24 and 1/32) for $\alpha=3$.
}
\end{figure}

\newpage

\setfigurenum{14}
\begin{figure}[ht]
\begin{center}
\includegraphics[scale=0.65]{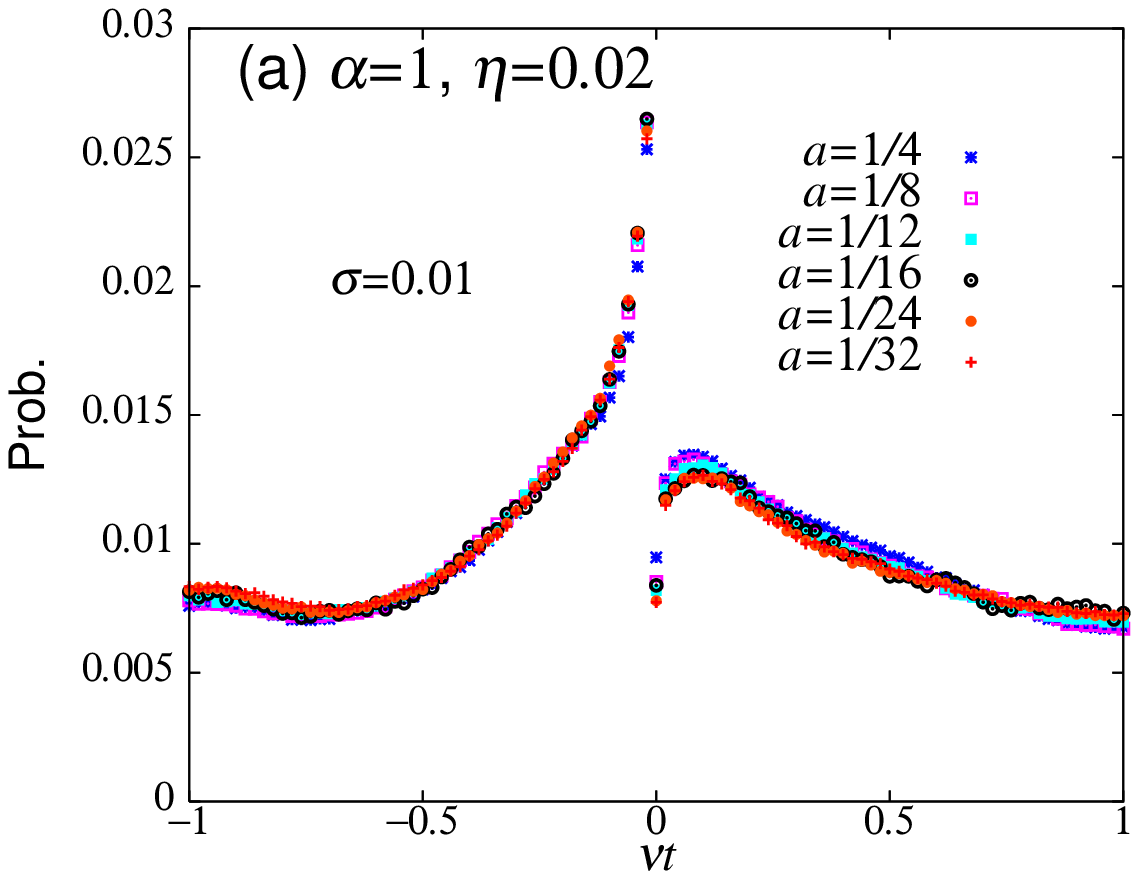}
\includegraphics[scale=0.65]{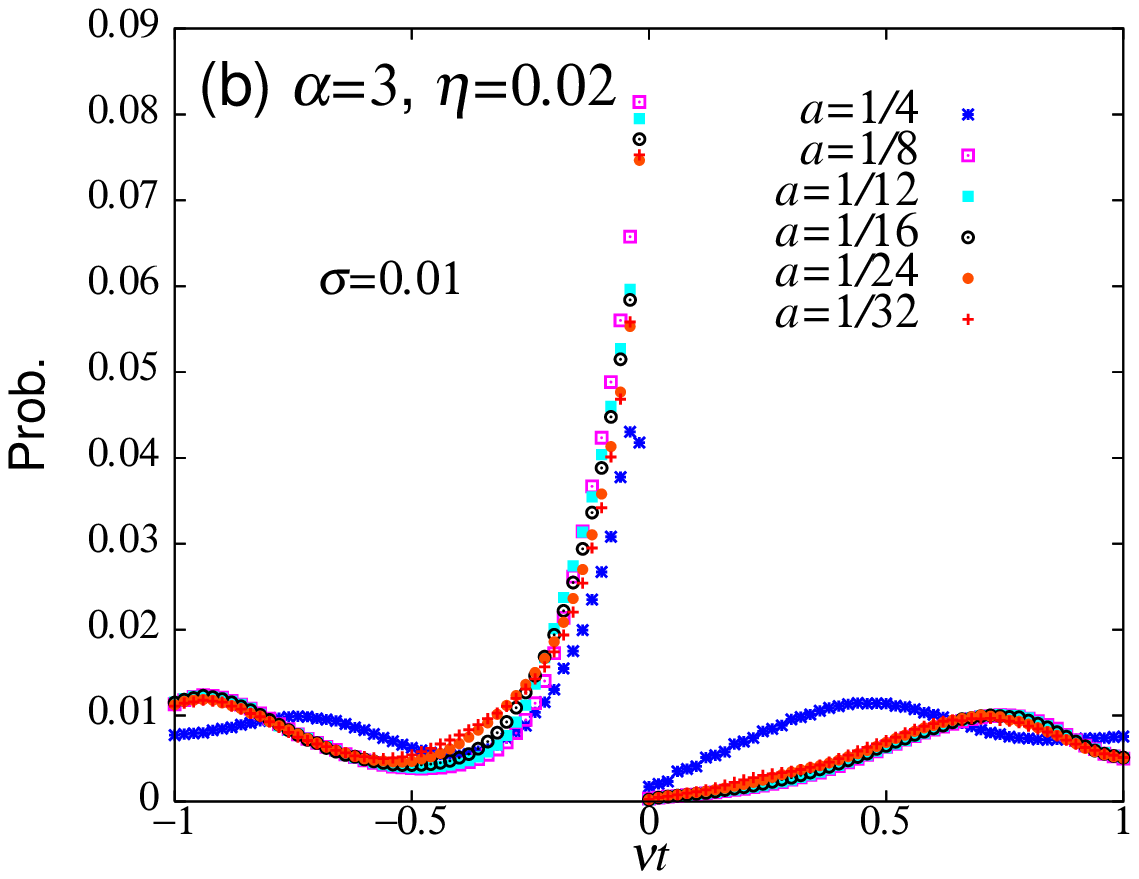}
\end{center}
\caption{
The time correlation function of the 1D viscosity BK model ($\eta=0.02$) between large events of $\mu_c=2$ (mainshock) occurring at time $t=0$ and
events of arbitrary sizes (dominated in number by small events) occurring at
time $t$. The dimensionless grid spacing  $a$ is varied in the range $1/4 \geq a \geq 1/32$. Fig. (a) represents the case of $\alpha=1$, while
Fig. (b) represents the case of $\alpha=3$. 
The parameter $\sigma$ is fixed to $\sigma =0.01$.
Events of arbitrary sizes occurring within the distance 
$r= 7.5$ from the epicenter of the mainshock are counted. 
The negative time $t<0$ represents the time before the mainshock, 
while the positive time $t>0$ represents the time after the mainshock. 
The average is taken over all large events of its magnitude 
$\mu >\mu_c=2$. The system size is $L=aN=200$. 
}
\end{figure}

\newpage

\setfigurenum{15}
\begin{figure}[ht]
\begin{center}
\includegraphics[scale=0.65]{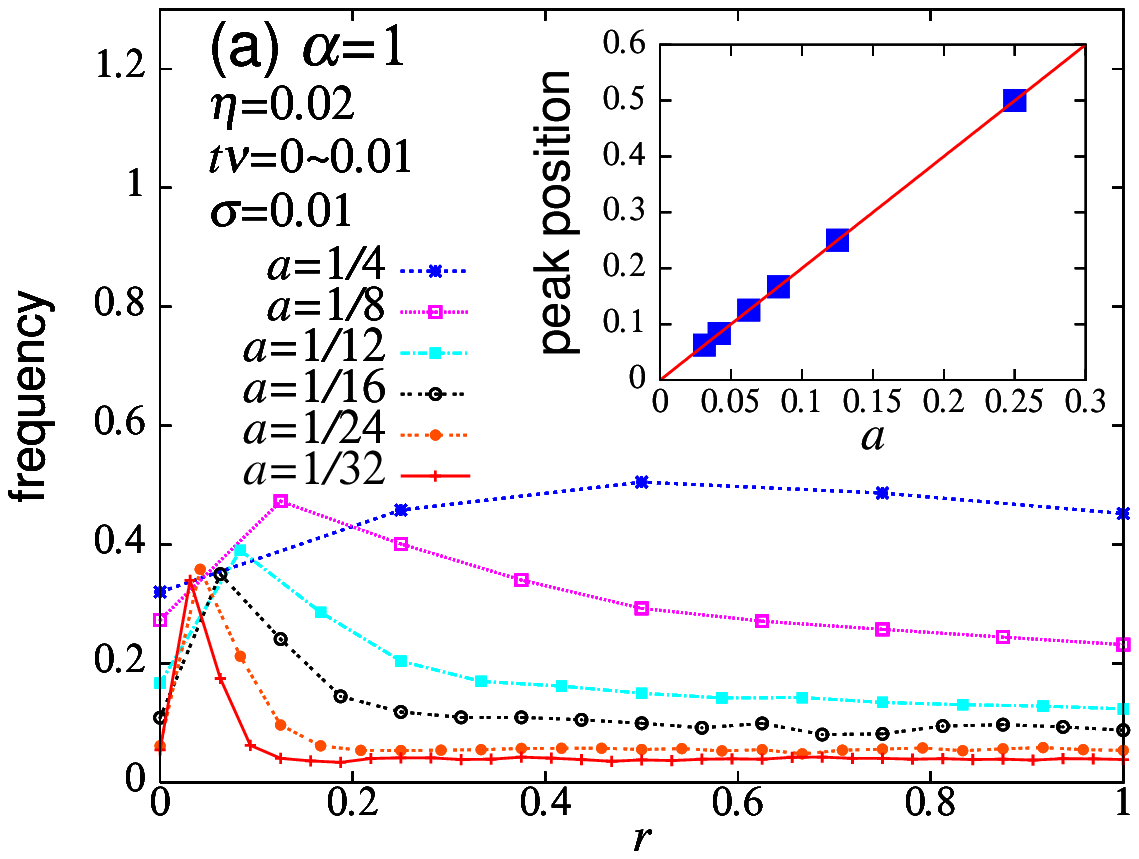}
\includegraphics[scale=0.65]{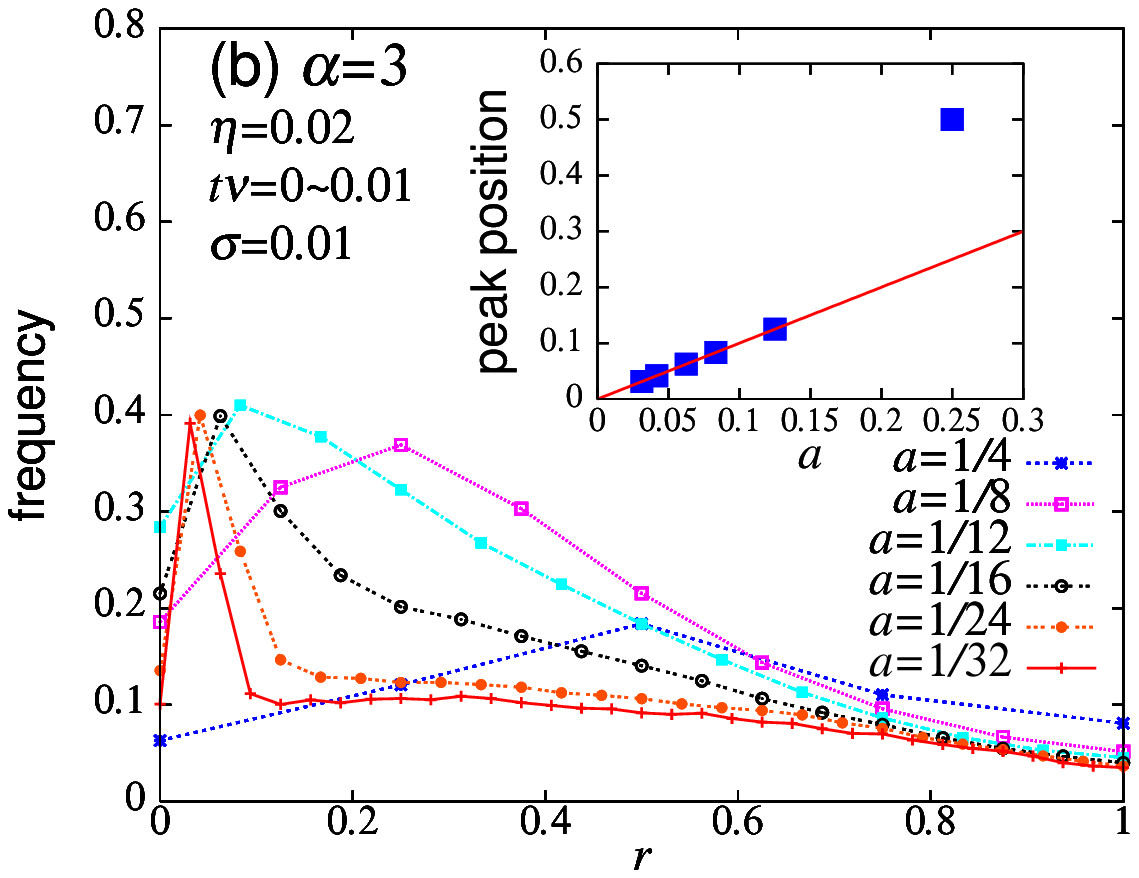}
\end{center}
\caption{
 Event frequency in the  time period $t\nu=0\sim0.01$ immediately before the mainshock of $\mu >\mu_c=2$  of the 1D viscosity BK model ($\eta=0.02$) plotted versus $r$, the distance from the epicenter of the upcoming mainshock. Fig. (a) represents the case of $\alpha=1$, while Fig. (b) represents the case of $\alpha=3$.  The dimensionless grid spacing $a$ is  varied in the range $1/4 \geq a \geq 1/32$. The parameter $\sigma$ is  fixed to $\sigma=0.01$. The system size is $L=aN=200$. The insets represent the peak position of the event frequency, corresponding to the range of the doughnut-like quiescence, as a function of the dimensionless grid spacing $a$. In both cases of $\alpha=1$ and 3, the doughnut-like quiescence appears to vanish in the continuum limit $a\rightarrow 0$.
}
\end{figure}

\newpage

\setfigurenum{16}
\begin{figure}[ht]
\begin{center}
\includegraphics[scale=0.65]{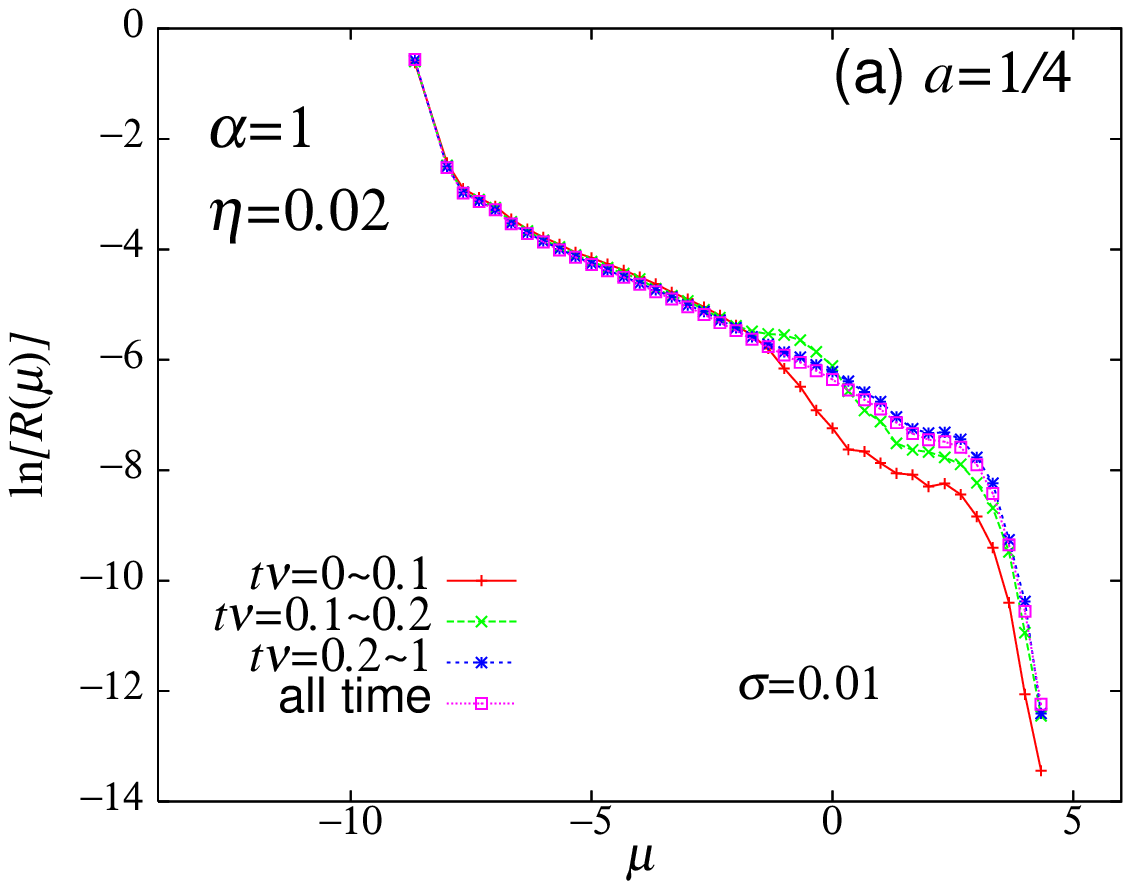}
\includegraphics[scale=0.65]{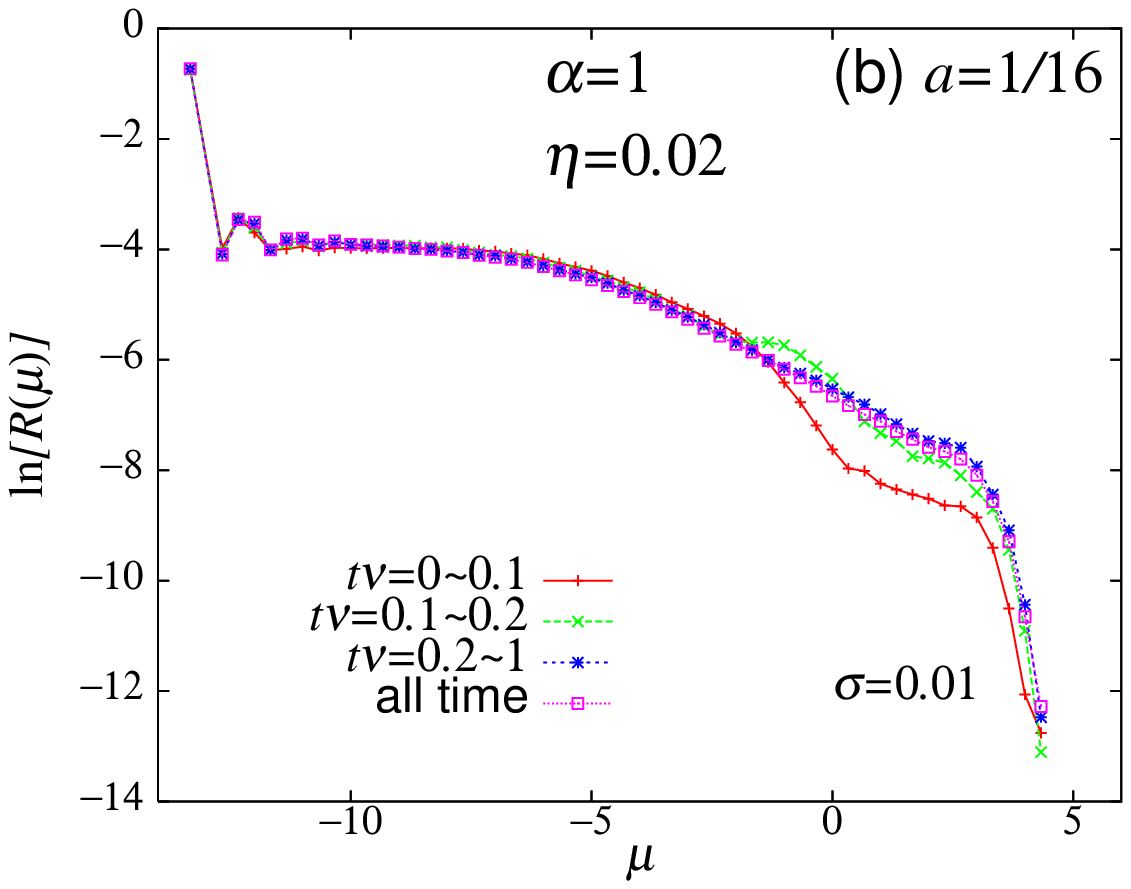}
\end{center}
\caption{
The local magnitude distribution for several time periods before the mainshock of $\mu >\mu _c=2$ of the 1D viscosity BK model 
($\eta=0.02$). The frictional parameters are $\alpha=1$ and  $\sigma=0.01$.  Figs. (a) and (b) represent the cases of $a=1/4$ (a) and $a=1/16$ (b), respectively. Events whose epicenter lies within the distance
$r=7.5$ from the epicenter of the upcoming mainshock are counted. The
system size is $L=aN=200$.
}
\end{figure}

\newpage

\setfigurenum{17}
\begin{figure}[ht]
\begin{center}
\includegraphics[scale=0.65]{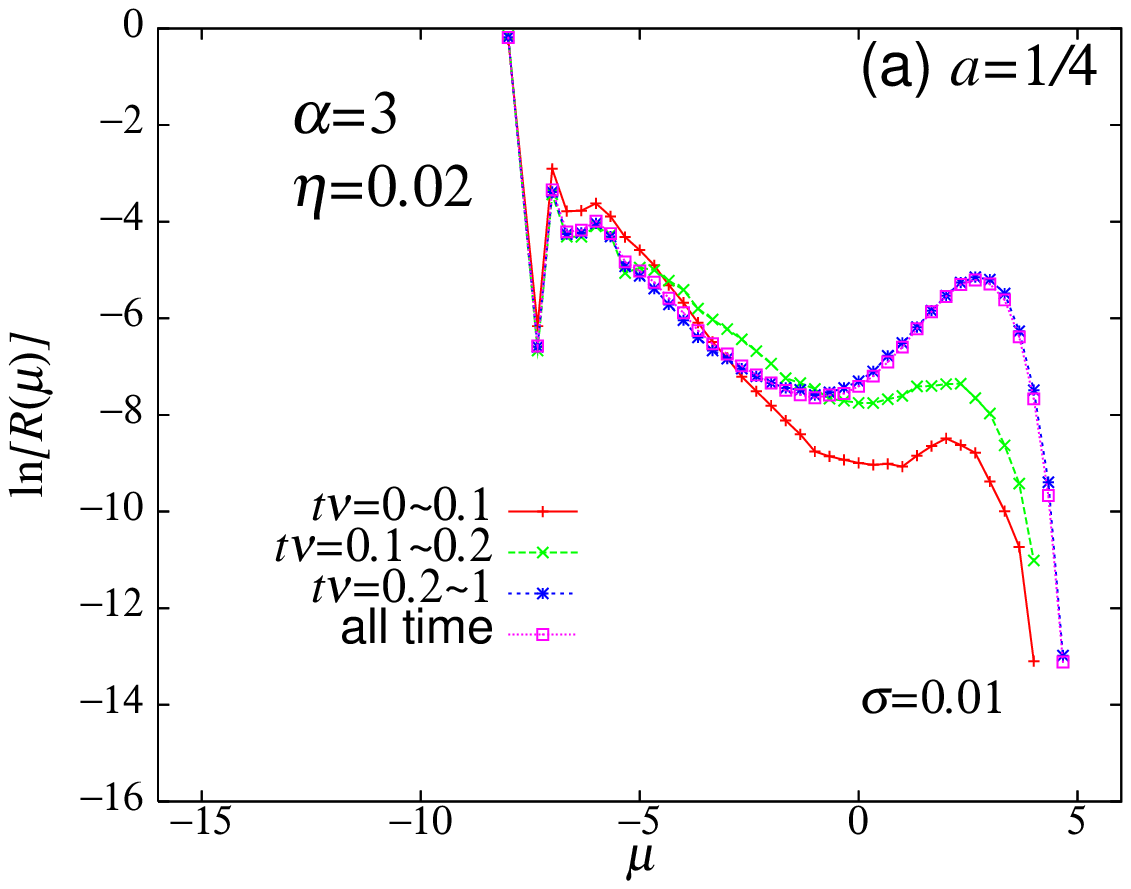}
\includegraphics[scale=0.65]{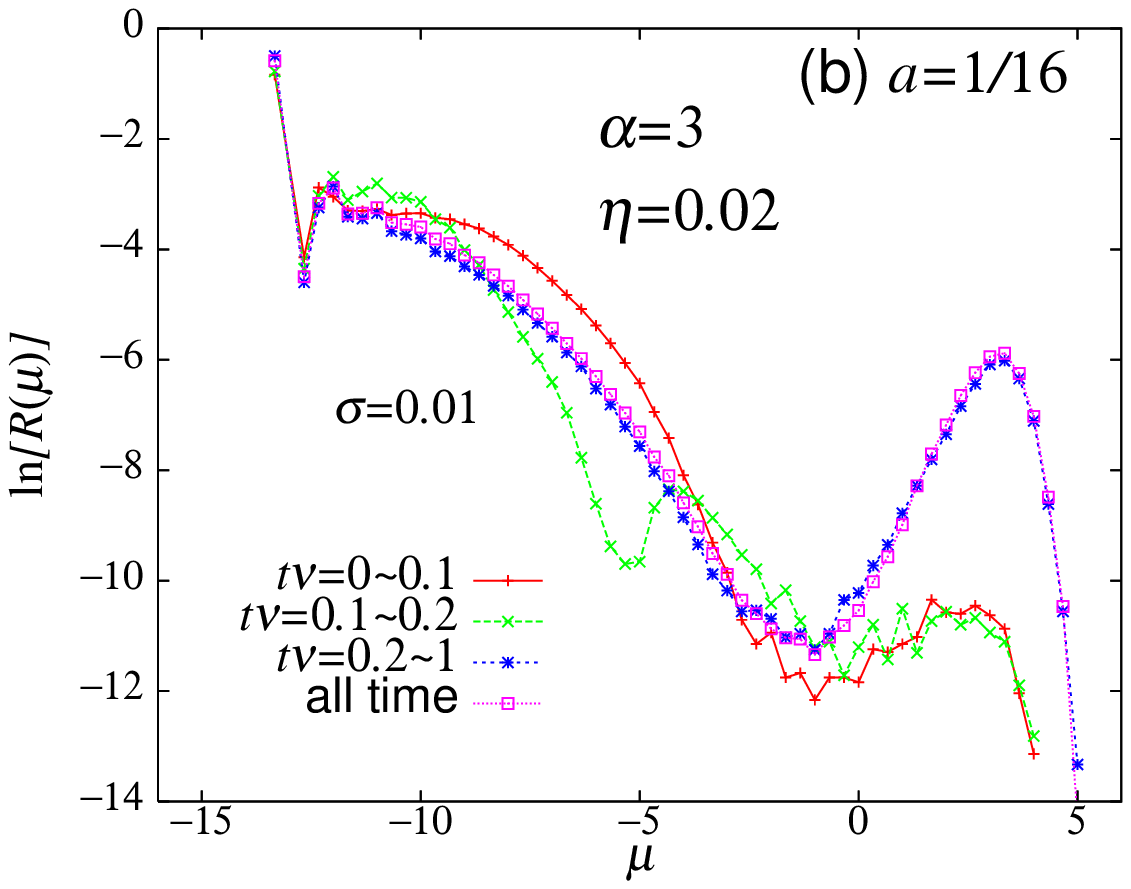}
\end{center}
\caption{
The local magnitude distribution for several time periods before the
mainshock of $\mu >\mu _c=2$ of the 1D viscosity BK model 
($\eta=0.02$). The frictional parameters are $\alpha=3$ and  $\sigma=0.01$. Figs. (a) and (b) represent the cases of $a=1/4$ (a) and $a=1/16$ (b), respectively.  Events whose epicenter lies within the distance
$r=7.5$ from the epicenter of the upcoming mainshock are counted. The
system size is $L=aN=200$. 
}
\end{figure}

\end{article}
\end{document}